\documentclass[twoside,twocolumn,9pt]{article}
\usepackage{extsizes}
\usepackage[super,sort&compress,comma]{natbib} 
\usepackage[version=3]{mhchem}
\usepackage[left=1.5cm, right=1.5cm, top=1.5cm, bottom=5.5cm]{geometry}
\usepackage{balance}
\usepackage{times,mathptmx}
\usepackage{sectsty}
\usepackage{graphicx} 
\usepackage{lastpage}
\usepackage[format=plain,justification=raggedright,singlelinecheck=false,font={stretch=1.125,small,sf},labelfont=bf,labelsep=space]{caption}
\usepackage{float}
\usepackage{fancyhdr}
\usepackage{fnpos}
\usepackage[english]{babel}
\usepackage{array}
\usepackage{charter}
\usepackage[T1]{fontenc}
\usepackage[usenames,dvipsnames]{xcolor}
\usepackage{setspace}
\usepackage[compact]{titlesec}
\newcommand{\fvec}{\ensuremath{\underline{f}}}
\newcommand{\qvec}{\ensuremath{\underline{q}}}

\newcommand{\rvec}{\ensuremath{\underline{r}}}
\newcommand{\rvecdot}{\ensuremath{\underline{\dot{r}}}}

\newcommand{\ddiff}{\ensuremath{\text{d}}}

\newcommand{\rl}{r_l}

\newcommand{\kB}{\mbox{$k_{\rm B}$}}
\newcommand{\kBT}{\mbox{$k_{\rm B}T$}}

\newcommand{\nsp}{\ensuremath{n_\mathrm{sp}}}

\newcommand{\muA}{\ensuremath{\mu_\mathrm{A}}}

\newcommand{\df}{d_{\rm f}}
\newcommand{\ds}{d_{\rm A}}

\newcommand{\smax}{s_\text{max}}

\newcommand{\Nend}{\ensuremath{N_\mathrm{e}}}
\newcommand{\mende}{\ensuremath{m_\mathrm{e+E}}}
\newcommand{\mendE}{\ensuremath{m_\mathrm{E}}}
\newcommand{\Rend}{\ensuremath{R_\mathrm{E}}}
\newcommand{\Rseg}{\ensuremath{R_\mathrm{s}}}

\newcommand{\Rgyr}{\ensuremath{R_\mathrm{g}}}

\newcommand{\wroots}{\ensuremath{w_{k=1}(s)}}
\newcommand{\Wwiener}{\ensuremath{W_\mathrm{1}}}
\newcommand{\rhohat}{\ensuremath{\hat{\rho}}}

\newcommand{\Nstar}{\ensuremath{N_{\star}}}

\newcommand{\cws}{\ensuremath{c_{\star}}}
\newcommand{\sone}{\ensuremath{s_1}}
\newcommand{\stwo}{\ensuremath{s_2}}
\newcommand{\none}{\ensuremath{n_1}}
\newcommand{\ntwo}{\ensuremath{n_2}}

\newcommand{\fone}{\ensuremath{f_1}}
\newcommand{\ftwo}{\ensuremath{f_2}}
\newcommand{\cone}{\ensuremath{c_\mathrm{asym}}}
\newcommand{\ctwo}{\ensuremath{c_\mathrm{corr}}}
\newcommand{\rcm}{\ensuremath{\rho_\mathrm{cm}}}
\newcommand{\rpair}{\ensuremath{\rho_\mathrm{pair}}}
\newcommand{\rcon}{\ensuremath{\rho_\mathrm{c}}}
\newcommand{\rconone}{\ensuremath{\rho_\mathrm{c,1}}}
\newcommand{\rcontwo}{\ensuremath{\rho_\mathrm{c,2}}}
\newcommand{\rconthree}{\ensuremath{\rho_\mathrm{c,3}}}
\newcommand{\rconlin}{\ensuremath{\rho_\mathrm{c,lin}}}

\newcommand{\lamp}{\ensuremath{\lambda_p}}
\newcommand{\taumon}{\ensuremath{\tau_\text{mon}}}
\newcommand{\taumax}{\ensuremath{\tau_\text{max}}}
\newcommand{\taup}{\ensuremath{\tau_p}}
\newcommand{\tauq}{\ensuremath{\tau_q}}
\newcommand{\taun}{\ensuremath{\tau_n}}
\newcommand{\tauN}{\ensuremath{\tau_N}}
\newcommand{\Rstar}{\ensuremath{R_{\star}}}

\newcommand{\Tstar}{\ensuremath{\tau_{\star}}}
\newcommand{\gmon}{\ensuremath{g_\text{mon}(t)}}
\newcommand{\gcms}{\ensuremath{g_\text{cm}(t)}}
\newcommand{\gcmsn}{\ensuremath{g_\text{cm,k}(t)}}
\newcommand{\Gamq}{\ensuremath{\Gamma_q}}

\definecolor{cream}{RGB}{222,217,201}

\begin{document}

\pagestyle{fancy}
\thispagestyle{plain}
\fancypagestyle{plain}{

%%%HEADER%%%
\fancyhead[C]{\includegraphics[width=18.5cm]{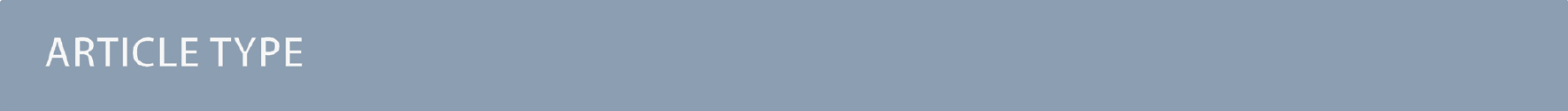}}
\fancyhead[L]{\hspace{0cm}\vspace{1.5cm}\includegraphics[height=30pt]{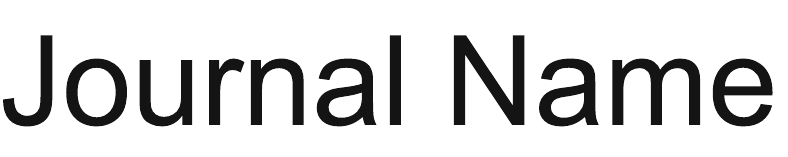}}
\fancyhead[R]{\hspace{0cm}\vspace{1.7cm}\includegraphics[height=55pt]{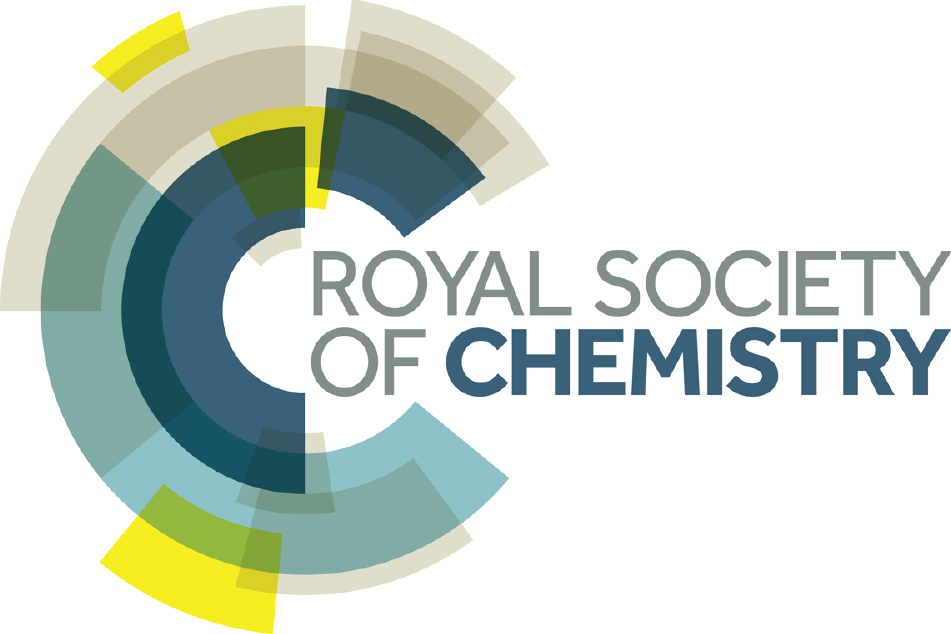}}
\renewcommand{\headrulewidth}{0pt}
}
%%%END OF HEADER%%%

%%%PAGE SETUP - Please do not change any commands within this section%%%
\makeFNbottom
\makeatletter
\renewcommand\LARGE{\@setfontsize\LARGE{15pt}{17}}
\renewcommand\Large{\@setfontsize\Large{12pt}{14}}
\renewcommand\large{\@setfontsize\large{10pt}{12}}
\renewcommand\footnotesize{\@setfontsize\footnotesize{7pt}{10}}
\makeatother

\renewcommand{\thefootnote}{\fnsymbol{footnote}}
\renewcommand\footnoterule{\vspace*{1pt}% 
\color{cream}\hrule width 3.5in height 0.4pt \color{black}\vspace*{5pt}} 
\setcounter{secnumdepth}{5}

\makeatletter 
\renewcommand\@biblabel[1]{#1}            
\renewcommand\@makefntext[1]% 
{\noindent\makebox[0pt][r]{\@thefnmark\,}#1}
\makeatother 
\renewcommand{\figurename}{\small{Fig.}~}
\sectionfont{\sffamily\Large}
\subsectionfont{\normalsize}
\subsubsectionfont{\bf}
\setstretch{1.125} %In particular, please do not alter this line.
\setlength{\skip\footins}{0.8cm}
\setlength{\footnotesep}{0.25cm}
\setlength{\jot}{10pt}
\titlespacing*{\section}{0pt}{4pt}{4pt}
\titlespacing*{\subsection}{0pt}{15pt}{1pt}
%%%END OF PAGE SETUP%%%

%%%FOOTER%%%
\fancyfoot{}
\fancyfoot[LO,RE]{\vspace{-7.1pt}\includegraphics[height=9pt]{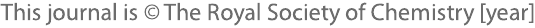}}
\fancyfoot[CO]{\vspace{-7.1pt}\hspace{13.2cm}\includegraphics{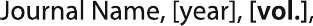}}
\fancyfoot[CE]{\vspace{-7.2pt}\hspace{-14.2cm}\includegraphics{RF}}
\fancyfoot[RO]{\footnotesize{\sffamily{1--\pageref{LastPage} ~\textbar  \hspace{2pt}\thepage}}}
\fancyfoot[LE]{\footnotesize{\sffamily{\thepage~\textbar\hspace{3.45cm} 1--\pageref{LastPage}}}}
\fancyhead{}
\renewcommand{\headrulewidth}{0pt} 
\renewcommand{\footrulewidth}{0pt}
\setlength{\arrayrulewidth}{1pt}
\setlength{\columnsep}{6.5mm}
\setlength\bibsep{1pt}
%%%END OF FOOTER%%%

%%%FIGURE SETUP - please do not change any commands within this section%%%
\makeatletter 
\newlength{\figrulesep} 
\setlength{\figrulesep}{0.5\textfloatsep} 

\newcommand{\topfigrule}{\vspace*{-1pt}% 
\noindent{\color{cream}\rule[-\figrulesep]{\columnwidth}{1.5pt}} }

\newcommand{\botfigrule}{\vspace*{-2pt}% 
\noindent{\color{cream}\rule[\figrulesep]{\columnwidth}{1.5pt}} }

\newcommand{\dblfigrule}{\vspace*{-1pt}% 
\noindent{\color{cream}\rule[-\figrulesep]{\textwidth}{1.5pt}} }

\makeatother
%%%END OF FIGURE SETUP%%%

%%%TITLE, AUTHORS AND ABSTRACT%%%
\twocolumn[
  \begin{@twocolumnfalse}
\vspace{3cm}
\sffamily
\begin{tabular}{m{4.5cm} p{13.5cm} }

\includegraphics{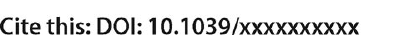} & \noindent\LARGE{\textbf{Marginally compact hyperbranched polymer trees}} \\
%Article title goes here instead of the text "This is the title"
\vspace{0.3cm} & \vspace{0.3cm} \\

 & \noindent\large{M.~Dolgushev,\textit{$^{a,b}$} J.P. Wittmer,$^{\ast}$\textit{$^{b}$} A. Johner,\textit{$^{b}$}
O. Benzerara,\textit{$^{b}$} H.~Meyer,\textit{$^{b}$} and J.~Baschnagel\textit{$^{b}$}} \\
%Author names go here instead of "Full name", etc.

\includegraphics{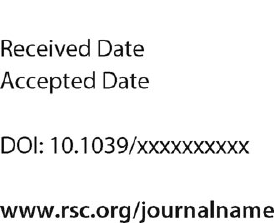} & \noindent\normalsize{Assuming Gaussian chain statistics 
along the chain contour, we generate by means of a proper fractal generator hyperbranched 
polymer trees which are marginally compact. Static and dynamical properties, such as the radial
intrachain pair density distribution $\rpair(r)$ or the shear-stress relaxation modulus $G(t)$, 
are investigated theoretically and by means of computer simulations.
We emphasize that  
albeit the self-contact density $\rcon = \rpair(r\approx 0) \sim \log(N/S)/\sqrt{S}$ diverges logarithmically with the
total mass $N$, this effect becomes rapidly irrelevant with increasing spacer length $S$.
In addition to this it is seen that the standard Rouse analysis must necessarily become inappropriate for compact objects for which
the relaxation time $\taup$ of mode $p$ must scale as $\taup \sim (N/p)^{5/3}$
rather than the usual square power law for linear chains.
} \\

\end{tabular}

 \end{@twocolumnfalse} \vspace{0.6cm}

  ]
%%%END OF TITLE, AUTHORS AND ABSTRACT%%%

%%%FONT SETUP - please do not change any commands within this section
\renewcommand*\rmdefault{bch}\normalfont\upshape
\rmfamily
\section*{}
\vspace{-1cm}

%%%FOOTNOTES%%%

\footnotetext{\textit{$^{a}$~Institute of Physics, University of Freiburg, Hermann-Herder-Str. 3, D-79104 Freiburg, Germany}}
\footnotetext{\textit{$^{b}$~Institut Charles Sadron, Universit\'e de Strasbourg \& CNRS, 23 rue du Loess, 67034 Strasbourg Cedex, France.}}
\footnotetext{\textit{$^{\ast}$ E-mail: joachim.wittmer@ics-cnrs.unistra.fr}}

%%%MAIN TEXT%%%%

\section{Introduction}
\label{sec_intro}

\paragraph*{Marginal compactness.}
%\paragraph*{Biological context.}
%
Natural selection quite generally has to strike a compromise between two 
requirements.\cite{FractalPhysiologyBook,WBE99}
On the one hand, biological structures have to be as compact (volume-filling) as possible
due to packing constraints and to reduce the typical spatial distances $R$ over which materials 
are transported within organisms and hence the time required for transport.
On the other hand, natural selection also tends to maximize the metabolic capacity of organs by 
increasing the average surface $A$ where resources are exchanged with the 
environment.\cite{MandelbrotBook,FractalPhysiologyBook,WBE99}
This leads to the extensive surface areas of, e.g., gills, lungs, guts, kidneys, sponges
and diverse respiratory and circulatory systems.
As a consequence of both tendencies a broad range of structures in biology form fractal networks 
\cite{MandelbrotBook,FractalPhysiologyBook,WBE99,Mirny11} 
which are, moreover, as {\em marginally compact} as possible.\cite{MSZ11,Mirny11}
This notion implies that the fractal bulk dimension $\df$ and the fractal surface dimension $\ds$ 
become similar approaching (from below) the dimension $d$ of the 
embedding space.\cite{MandelbrotBook} (It is assumed below that $d=3$.)
The average linear size $R(N)$ --- characterized, e.g., by the radius of gyration $\Rgyr(N)$ ---
and the surface $A(N)$ --- obtained, e.g., from the number of subunits interacting physiologically with the 
environment --- thus increase in the large-$N$ limit as
\begin{eqnarray}
R(N) & \sim & N^{1/\df} \approx N^{1/d} \mbox{ and } \label{eq_df} \\
A(N) & \sim & R^{\ds} \sim N^{\ds/\df} \approx N, \label{eq_ds}
\end{eqnarray}
i.e. every subunit has thus to leading order the same $N$-independent finite probability 
to interact with the environment.\footnote[2]{All subunits of {\em open} objects ($\df < d$) 
are at or close to the surface, i.e. $A(N) \sim N$. If one insists on using the first relation 
of eqn~(\ref{eq_ds}) as the operational definition of $\ds$, this implies $\ds=\df$ for open objects.
The more common (and perhaps mathematically more rigorous) definition
of the surface fractal dimension $\ds$ assumes that the considered object
is compact ($\df \equiv d$).\cite{MandelbrotBook}}
Importantly, marginal compactness is observable experimentally from the scaling 
of the structure (form) factor $F(q)$, i.e. the Fourier transformed pair-correlation function of the 
relevant subunits of the network under consideration 
($q$ being the wavevector).\cite{MandelbrotBook,RubinsteinBook,BenoitBook}
As reminded in Section~\ref{stat_Fq}, it can be shown\cite{MSZ11} that to leading order we have 
\begin{equation}
F(q) \approx N/[R(N)q]^d \sim N^0 \mbox{ for } 1/R(N) \ll  q \ll 1/b
\label{eq_Fq_margcomp}
\end{equation}
with $b$ being a local length scale (lower cutoff) 
which is often set by the size of the subunits of the network.

\paragraph*{A controversial example.}
Following a first brief comment by some of us,\cite{MSZ11} there appears to be a growing 
(albeit not general) consensus \cite{MSZ11,WMJ13,obukhovmodel,Rubinstein16,Grosberg16,Michieletto16}
that unknotted and unconcatenated polymer rings in dense solutions and melts 
may reveal a similar marginally compact behavior. % with $R(n) \sim n^{1/3}$ and $A(n) \sim n$.
That such rings should adopt increasingly compact configurations has been expected 
theoretically due to the mutual repulsion caused by the topological 
constraints.\cite{Cates86,Obukhov94,KN96,Rubinstein16}
Various numerical 
\cite{MWC96,MWC00,KK11a,KK11b,KK13,MSZ11,WMJ13,obukhovmodel,Everaers14,Turner16} 
and experimental studies \cite{Rubinstein08b,Rubinstein13,Goossen14,Goossen15} 
suggest that the apparent fractal dimension $\df(N)$ approaches $d=3$ with increasing mass $N$. 
Naturally, this begs the question of how to characterize the surface of these assumed ultimately compact objects.
Since there is no obvious reason for a finite surface tension, a non-Euclidean irregular surface
is expected to be characterized by an apriori unknown fractal surface dimension with $2 < \ds \le d$.
The limit $\ds \to d^{-}$ is an attractive scenario 
since all monomers are evenly exposed to the topological constraints imposed by other chains, 
i.e. all subchains have the same self-similar and isotropic statistics 
(no screening of topological interactions).
Interestingly, motivated by the behavior of melts of strictly two-dimensional linear chains,\cite{ANS03}
a different hypothesis has been suggested in the recent molecular dynamics (MD) simulations
and numerical analysis of ring melts.\cite{KK11a,KK11b,KK13,KK13b}
Various properties are fitted with power-law exponents corresponding to (in our language) fractal surface 
dimensions $\ds$ similar to $d$. The reported differences are, however, too small ---
considering the limited number of decades available at present
and that error bars must be interpreted with care ---
to rule out marginal compactness merely on {\em numerical} grounds.
Much more important is the clever {\em theoretical} argument\cite{KK11a}
that marginal compactness implies that the return probability $p(n)$ 
between two tagged monomers must decay inversely with the mass $n$ of the chain 
segment.\footnote[3]{While for rings $n$ is equivalent to the arc-length $s$ along the chain contour,
this notion must be generalized for branched structures as discussed in Section~\ref{stat_pn}.
Please note that our definition of $n$ is slightly different from the one used
in recent work on rings and branched polymers.\cite{Everaers16a,Everaers16b,Everaers17,Everaers17b,Grosberg14,Grosberg16}}
Since this yields in turn a logarithmically diverging self-contact density 
\begin{equation}
\rcon \equiv \sum_{n=1}^N p(n) \sim \log(N),
\label{eq_rcon_log}
\end{equation}
it is argued that, quite generally,
a ``mathematically rigorous fractal structure" can not be marginally compact.\cite{KK11a}
\begin{figure}[t]
\centerline{\resizebox{0.9\columnwidth}{!}{\includegraphics*{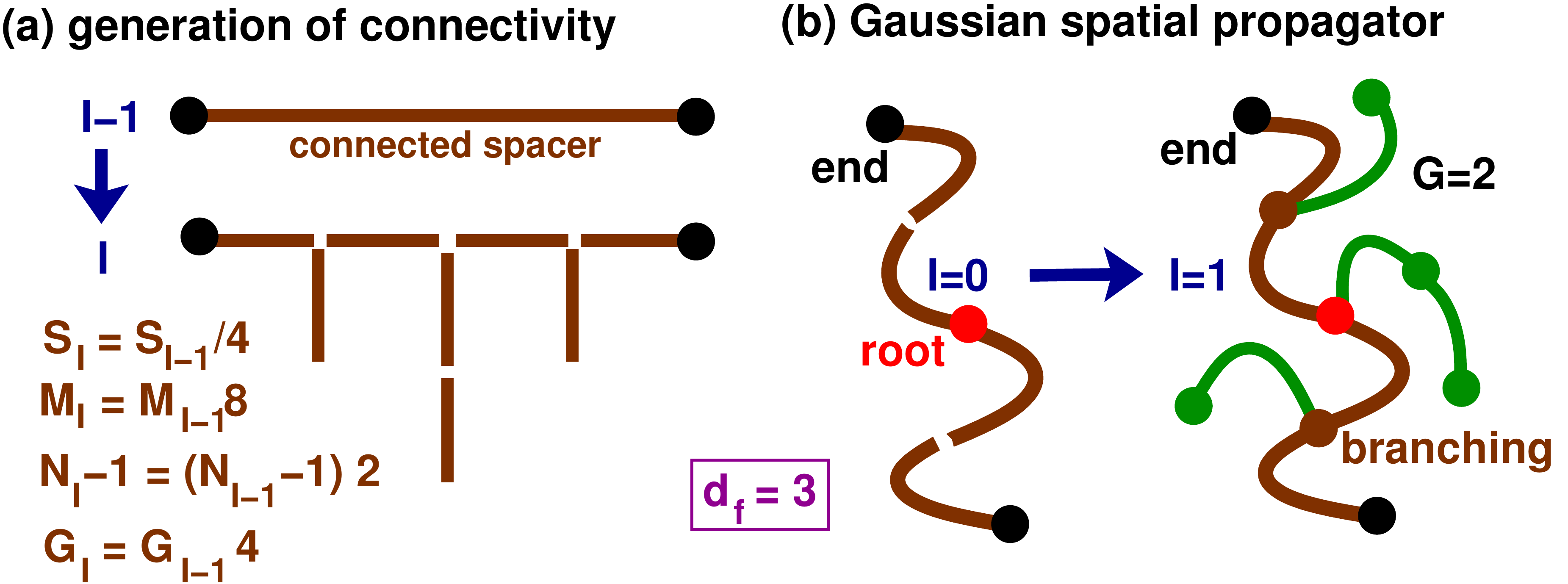}}}
\caption{Iterative generation of marginally compact hyperbranched trees in $d=3$ dimensions
by means of a proper fractal generator:
{\bf (a)} 
A given segment of curvilinear length $S_{I-1}$ is replaced by $M_I = 8 M_{I-1}=2^{3I}$
segments of length $S_I=S_{I-1}/4 = S_0/2^{2I}$ with $I$ denoting the number of iterations.
{\bf (b)} 
While the connectivity matrix is generated in a deterministic manner,
the particle positions are chosen randomly assuming Gaussian chain statistics along the chain contours
as shown for the iteration $I=0 \to I=1$. 
}
\label{fig_sketch}
\end{figure}

\paragraph*{Central goal of the current study.}
Not focusing specifically on melts of rings, but addressing this argument for general marginally
compact structures, we want to show in the present work that there is a simple, generic way out of this difficulty.
As sketched in Fig.~\ref{fig_sketch}, we investigate as a counter example a marginally compact hyperbranched tree 
generated using {\em one} proper fractal generator.
The present study elaborates on one of the models introduced in a recent work on dendrimers and 
hyperbranched stars with Gaussian chain statistics along the chain contour.\cite{dendgauss}
%
%A similar, more complicated two-generator (multi-fractal) model has been proposed 
%to characterize the sluggish crossover for rings.\cite{obukhovmodel}
%
%The simpler one-generator model has the advantage (for our purpose) that the 
%trees become self-similar marginally compact objects after two iterations of the generator.
%
As shown by our toy model, it is straightforward to generate {\em mathematically} rigorous marginally compact objects.
They simply do exist and this irrespectively of whether some moments of $p(n)$ diverge with increasing mass $N$.
It is another issue, however, whether for the real system, one attempts to describe using the mathematical model,
there exist additional physical or biological requirements which set constraints on moments of $p(n)$.
Quite generally, such a constraint may lead to a restriction of the parameter space within
which the fractal model may be applied and may set an upper bound $\Nstar$ for the system size $N$.
The existence of an upper bound does not invalidate {\em per se} a fractal description,
at least not if a broad $N$-window can be identified where the constraints are irrelevant.
The {\em physical} (not mathematical) restriction we need to address is the fact that the number 
of close contacts of any reference monomer is generally limited due to excluded volume constraints.\footnote[4]{Hyperbranched 
trees with fractal dimension $\df=4$ have been discussed, e.g., in the context of randomly branched polymers 
(often called ``lattice animals"),\cite{Stock49} dilute rings in a gel of topological obstacles \cite{Obukhov94} and
as an early (mathematically perfectly legitimate) model describing the topological
interactions of unconcatenated molten rings.\cite{Obukhov94,Rubinstein08b,Rubinstein13}
Obviously, such a modeling approach must break down in $d=3$ dimension
above an upper mass limit $\Nstar$.
A sufficiently broad mass window exists, however, where the average self-density
remains below unity and the fractal model is thus applicable.\cite{Obukhov94}
Please note that the limit set by the local excluded volume constraint on the randomly branched
lattice animals is much more severe as the one, eqn~(\ref{eq_keyone}),
set on our marginally compact trees.}
%\cite{foot_lattanimal}.
%
We show below that albeit the monomer connectivity in a marginal compact tree leads
indeed to a logarithmically diverging self-contact density, in agreement with eqn~(\ref{eq_rcon_log}), 
there exist an extremely broad $N$-window
\begin{equation}
%\rcon \sim \log(N/S)/\sqrt{S},
N \ll \Nstar \approx S e^{\sqrt{S}},
\label{eq_keyone}
\end{equation}
strongly increasing with the spacer length $S$ between the branching points of network,
where the excluded volume constraints can be neglected.
(To make the above exponential relation meaningful, prefactors must be specified.
This will be done in Section~\ref{stat_contact}.)
We thus argue that one cannot reject {\em apriori} %--- neither on mathematical nor on physical grounds ---
a marginally compact model such as the one proposed by Obukhov {\em et al}.\cite{obukhovmodel}

\paragraph*{Outline.}
The fractal generator and the construction of the hyperbranched trees is described in Section~\ref{sec_model}
where we also give some algorithmic details concerning the MD and Monte Carlo (MC) simulations
\cite{AllenTildesleyBook} used in the present work.
Static properties are then presented in Section~\ref{sec_stat} where we confirm 
eqn~(\ref{eq_Fq_margcomp}) and eqn~(\ref{eq_keyone}).
%
%The true structure factor $F(q)$ and the Fourier transformation of the average density 
%profile are shown in Section~\ref{stat_Fq} to be rather different since density fluctuations 
%matter albeit our trees are compact. 
%
We turn then in Section~\ref{sec_dyna} to dynamical properties. 
%The standard linear-chain Rouse scaling 
%\cite{RubinsteinBook,DoiEdwardsBook} is shown to be inappropriate as one expects due to the 
%compactness of the chain on all scales. 
%
The work is summarized in Section~\ref{sec_conc}.
Appendix~\ref{app_contact} gives some details concerning the prefactors expected for
the self-contact density $\rcon$. A comparison of the structure factor of our model
and (already published) simulated ring data\cite{MSZ11} is given in Appendix~\ref{app_Fq}. 
Some properties of the ``generalized Rouse model" (GRM) 
\cite{Blumen03,dolg15,dolg16a,dolg16b,dolg16c,dolg16d,Biswas10,Biswas12}
are recalled in Appendix~\ref{app_theo}.
%
%A technical point concerning the shear-stress relaxation modulus $G(t)$ is made in Appendix~\ref{app_Gt}.
%%%%%%%%%%%%%%%%%%%%%%%%%%%%%%%%%%%%%%%%%%%%%%%%%%%%%%%%%%%
\section{Model}
\label{sec_model}
\begin{figure}[t]
\centerline{\resizebox{0.7\columnwidth}{!}{\includegraphics*{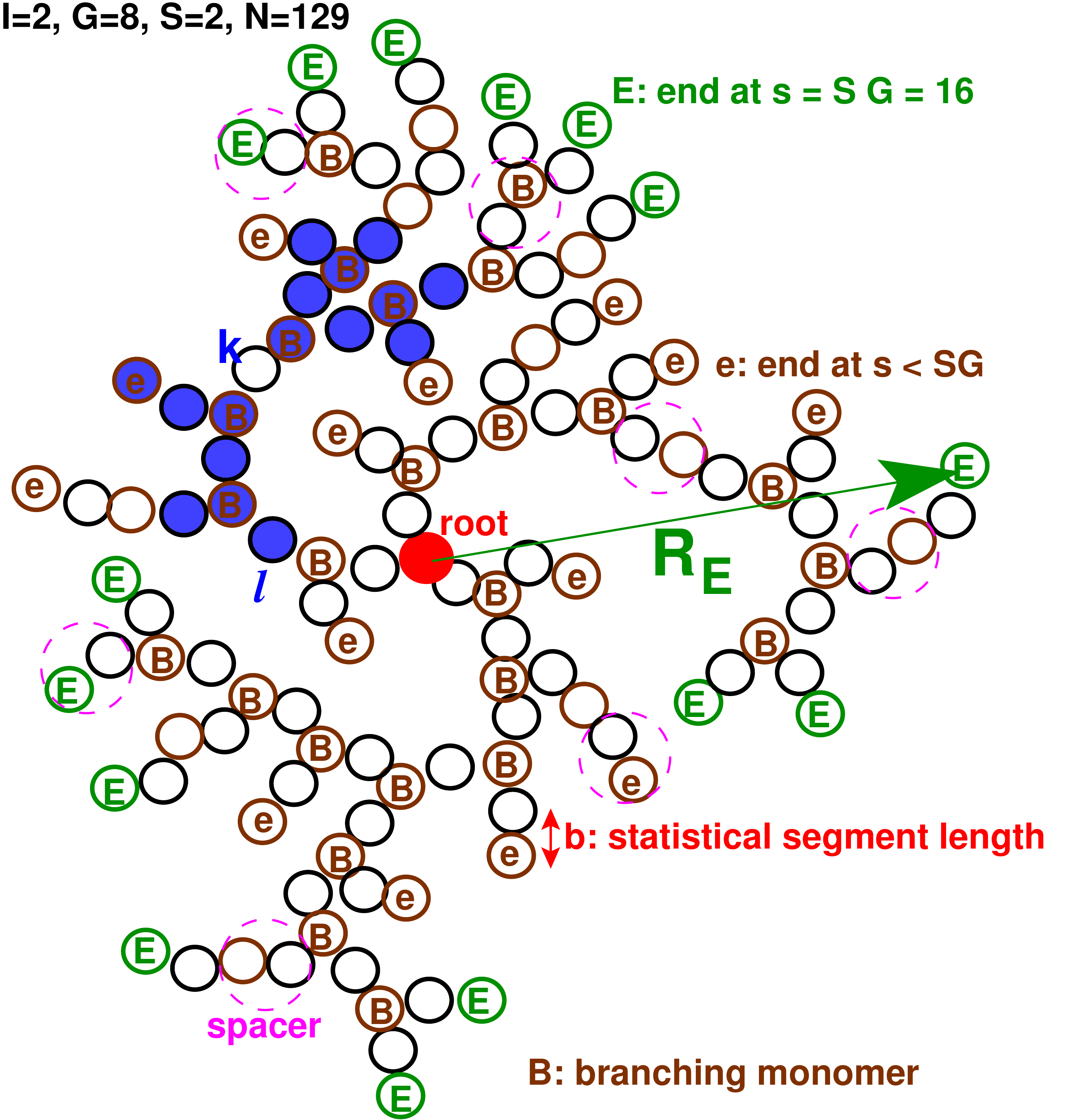}}}
\caption{Sketch of some notations using an example with iteration $I=2$, number of generations $G=8$, 
spacer length $S=2$, total monomer number $N=129$ and total number of dangling ends $\mende=29$. 
Some spacer chains of lengths $S=2$ are marked by dashed circles,
branching points by ``B", dangling ends at largest curvilinear distance $S G$ from the root monomer by ``E", 
all other dangling ends by ``e".  $\Rend$ denotes the root-mean-squared distance between the root monomer and the ends ``E".
The $n(s)$ monomers with a curvilinear distance $\le s=4$ from the monomer $k$ are marked by filled disks.
}
\label{fig_sketch_detail}
\end{figure}

\paragraph*{Gaussian chain statistics.}
Treelike networks with Gaussian chain statistics have been considered theoretically 
early in the literature \cite{Stock49,Burchard82,Obukhov94,Hammouda92,Biswas94} 
and have continued to attract attention up to the recent 
past.\cite{Blumen03,dolg15,dolg16a,dolg16b,dolg16c,dolg16d,Biswas10,Biswas12,Nechaev15}
Assuming translational invariance, these models have in common that
the root-mean-square distance $\Rseg$ between two monomers $i$ and $j$ is given by
$\Rseg^2 \equiv \langle (\rvec_i - \rvec_j)^2 \rangle = b^2 s$
with $\rvec_i$ being the position of a bead $i$, 
$s$ the shortest chemical distance (curvilinear distance) connecting both 
monomers and $b$ the statistical segment size.\cite{DoiEdwardsBook}
Other moments are obtained from the distribution $P(\rvec,s)$ of the distance vector $\rvec = \rvec_i-\rvec_j$ 
which, irrespective of the specific topology, is given by 
\begin{equation}
P(\rvec,s) = 
\left(\frac{d}{2\pi \Rseg^2}\right)^{d/2} \exp\left(-\frac{d}{2} \left(\frac{r}{\Rseg} \right)^2 \right)
\label{eq_Grsgauss}
\end{equation}
with $d=3$ being the spatial dimension. 
Such Gaussian trees may be modeled theoretically as well as in MD and MC simulations
using $\nsp$ ideal harmonic springs connecting specified pairs of beads $i$ and $j$. 
The potential energy $V$ may be written as
\begin{equation}
V = \frac{K}{2} \sum_{l=1}^{\nsp} \rl^2 
= \frac{K}{2} \sum_{i,j=1}^N A_{ij} \  \rvec_i \cdot \rvec_j 
\label{eq_Vspring}
\end{equation}
with $K= d \kBT/b^2$ being the spring constant,
$l$ a label standing for the spring between the monomers $i$ and $j$,
$\rl=|\rvec_i-\rvec_j|$ the corresponding monomer distance and 
$\mathbf{A}=(A_{ij})$ the so-called connectivity or 
Laplacian matrix.\cite{Blumen01,Biswas14,Biswas16}
This is a $N \times N$ square matrix with diagonal elements $A_{ii}$ denoting the number of bonds 
connected to monomer $i$ and off-diagonal elements $A_{ij}=-1$ if the beads $i$ and $j$ are 
connected or $A_{ij}=0$ elsewise.
Due to their theoretical simplicity such Gaussian chain trees 
(including systems with {\em short-range} interactions along the network)
allow to investigate non-trivial conceptual and technical issues, both for 
static \cite{dolg15,dolg16c,dendgauss} and dynamical 
\cite{Biswas94,Mendoza06,Wu12,Blumen01,dolg16a,dolg16b,dolg16c,dolg16d,Biswas14,Biswas16} properties. 

\paragraph*{Construction of connectivity matrix.}
It is assumed in the present study that the hyperbranched network is not annealed, 
but irreversibly imposed by the fractal generator sketched in Fig.~\ref{fig_sketch}.
Note that fractal self-similar hyperbranched trees generated in this way are called ``$\beta$-stars"
in \citet{dendgauss} where a more detailed description can be found.
Basically, we start at iteration step $I=0$ with a linear Gaussian chain of $1+S_0$ monomers created using 
the distribution eqn~(\ref{eq_Grsgauss}).
The ``$1$'' is required due to the ``root monomer" sitting for all iterations $I$ 
topologically in the middle of the network. 
This was done for technical convenience to have a fixed reference monomer for the various pointer lists used.
(The iteration step $I=0\to I=1$ is thus slightly different from all others.)
At each iteration step $I-1 \to I$ each spacer subchain of length $S_{I-1}$ is divided into 4 subchains of 
equal length $S_I=S_{I-1}/4=S_0/2^{2I}$ and 4 subchains of same length are added laterally to the subchain 
as sketched in Fig.~\ref{fig_sketch}. 
The total number of spacer chains $M_I$ thus increases as
$M_I = 8 M_{I-1} = 2^{3I}$ and the total mass $N_I$ as $N_I-1 = 2 (N_{I-1}-1) = 2^I S_0 = M_I S_I$. 
Note that the 4 new spacer subchains are again created according to eqn~(\ref{eq_Grsgauss}).
Importantly, $S_0$ may be chosen such that for the final iteration $I$ considered 
we have the same monodisperse final spacer length $S\equiv S_I$. Below we shall often use
$N \equiv N_I$ for the total mass after $I$ iterations, $\smax \equiv S_0 = 2^{2I}S$ for the
largest possible curvilinear distance and $G=\smax/2S=2^{2I-1}$ for the number of 
``generations" of spacer chains of length $S$. We thus have by construction
\begin{equation}
N-1 = 2^{dI}S = S (\smax/S)^{d/2} = S (2G)^{d/2}
\label{eq_Ntot}
\end{equation}
with $d=3$.\footnote[5]{Note that the exponent $d/2$ corresponds to the exponent $1/\rho$ used
in various recent publications on melts of ring polymers and annealed branched 
polymers.\cite{Everaers16a,Everaers16b,Everaers17,Everaers17b,Grosberg14,Grosberg16}
At difference to these studies $1/\rho$ is imposed here, not fitted or predicted.} 
As a consequence the typical distance $\Rend$ between the root monomer and the monomers ``E" (Fig.~\ref{fig_sketch_detail}), 
as one observable measuring the tree size, must scale as $\Rend = b \sqrt{S G} \sim N^{1/d}$.

\begin{table*}[t]
\begin{center}
\begin{tabular}{cccccccccc}
\hline
$I$&$G$ &$N$   &$\mende$&$\mendE$&$\smax$&$\Wwiener/N^2$&$\Rend$&$\Rgyr$ &$\rcon$\\ \hline
1  &2   &9     &5       & 5      &4      &1.2     &1.41   &1.09 &1.01   \\
2  &8   &65    &29      & 15      &16     &4.2     &2.82   &2.06 &2.06  \\
3  &32  &513   &221     & 45      &64     &16      &5.66   &4.00 &3.18  \\
4  &128 &4097  &1757    & 135      &256    &63      &11.3   &7.95 &4.34  \\
5  &512 &32769 &14045   & 405     &1024   &252     &22.6   &15.9 &5.52  \\
6  &2048&262145&112349  & 1215     &4096   &1008    &45.3   &31.4 &6.71  \\
\hline
\end{tabular}
\vspace*{0.5cm}
\caption[]{Some properties for spacer length $S=1$:
iteration number $I$,
generation number $G$,
total mass $N$,
total number of dangling ends $\mende$,
number of ends $\mendE = 5 \times 3^{I-1}$ with maximum distances $S G$ from root monomer, 
maximum possible curvilinear arc-length $\smax$,
rescaled Wiener index $\Wwiener/N^2$ (Section~\ref{stat_ws}), 
root-mean-squared distance $\Rend$ between root monomer and dangling end ``E",
radius of gyration $\Rgyr$ (Section~\ref{stat_compact}) and 
self-contact density $\rcon$ (Section~\ref{stat_contact}).
\label{tab_I}}
\end{center}
\end{table*}

\begin{table}[t]
%\begin{center}
\begin{tabular}{cccccccc}
\hline
$S$&$N-1$   &$\smax$ &$\Wwiener/N^2$&$\Rend$&$\Rgyr$ &$N/\Rgyr^3$ &$\rcon$\\ \hline
1  &$2^{15}$&1024    &252.2   &22.6   &15.9 &8.19 &5.52  \\
2  &$2^{16}$&2048    &503.9   &32     &22.5 &5.79 &4.59  \\
4  &$2^{17}$&4096    &1007.5  &45.3   &31.7 &4.09 &3.85  \\
8  &$2^{18}$&8192    &2014.5  &64     &44.5 &2.89 &3.29  \\
16 &$2^{19}$&$2^{14}$&4028    &90.5   &71.5 &2.05 &2.85  \\
32 &$2^{20}$&$2^{15}$&8056    &128    &89.8 &1.45 &2.54  \\
64 &$2^{21}$&$2^{16}$&16112   &181    &126.9&1.03 &2.31  \\
128&$2^{22}$&$2^{17}$&32224   &256    &179.5&0.73 &2.14  \\
256&$2^{23}$&$2^{18}$&64448   &362    &253.9&0.51 &2.02  \\
512&$2^{24}$&$2^{19}$&123274  &512    &359  &0.36 &1.94  \\
\hline
\end{tabular}
\vspace*{0.5cm}
\caption[]{Some properties for iteration $I=5$:
spacer length $S$,
total mass $N-1$,
maximum arc-length $\smax$,
Wiener index $\Wwiener/N^2$, 
root-mean- squared distance $\Rend$, radius of gyration $\Rgyr$,
$N/\Rgyr^3$ as a measure of the typical density 
and self-contact density $\rcon$.
Note that $N/\Rgyr^3 \sim 1/\sqrt{S}$ and that $\rcon$ converges similarly
down to $\rconlin = 1.724/b^3$ for large $S$.
\label{tab_S}}
%\end{center}
\end{table}

\paragraph*{Technical comments.}
Due to their Gaussian chain statistics many conformational properties can be readily obtained 
using Gaussian propagator techniques or equivalent linear algebra 
relations.\cite{dolg15,dolg16d,Biswas10,Biswas11,Biswas12}
However, some interesting properties, 
such as the dynamical structure factor $F(q,t)$ discussed in Section~\ref{dyna_Fqt},
can be more easily computed by direct simulation which are in any case necessary if long-range 
interactions between the monomers are switched on. 
Following \citet{dendgauss} we have used for static properties MC simulations
mixing pivot and local jump moves. To obtain dynamical properties we have in addition 
sampled time series using local MC moves with maximum jump distance
$\delta r =0.5$ and velocity-verlet MD using a Langevin thermostat \cite{AllenTildesleyBook,LAMMPS}
with an extremely strong friction constant $\zeta =10$ suppressing {\em all} inertia effects. 
(This is of course not the most efficient parameter to sample the configuration space via MD.)
The temperature $T$, Boltzmann's constant $\kB$ and the statistical bond length $b$ are set to unity. 
We sample in parallel $m=100$ uncorrelated chains. These are contained in a periodic cubic 
simulation box of total density $\rho = N m/V = 1$ with $V$ being the volume. 
We focus often on trees with a spacer length $S=1$ as summarized in Table~\ref{tab_I}.
As may be seen from Table~\ref{tab_S} for $I=5$, a broad range of spacer lengths $S$ has also been sampled.
This is especially important for the scaling of the self-contact density $\rcon$ presented in Section~\ref{stat_contact}. 

%%%%%%%%%%%%%%%%%%%%%%%%%%%%%%%%%%%%%%%%%%%%%%%%%%%%%%%%%%%
\section{Static properties}
\label{sec_stat}
\begin{figure}[t]
\centerline{\resizebox{0.9\columnwidth}{!}{\includegraphics*{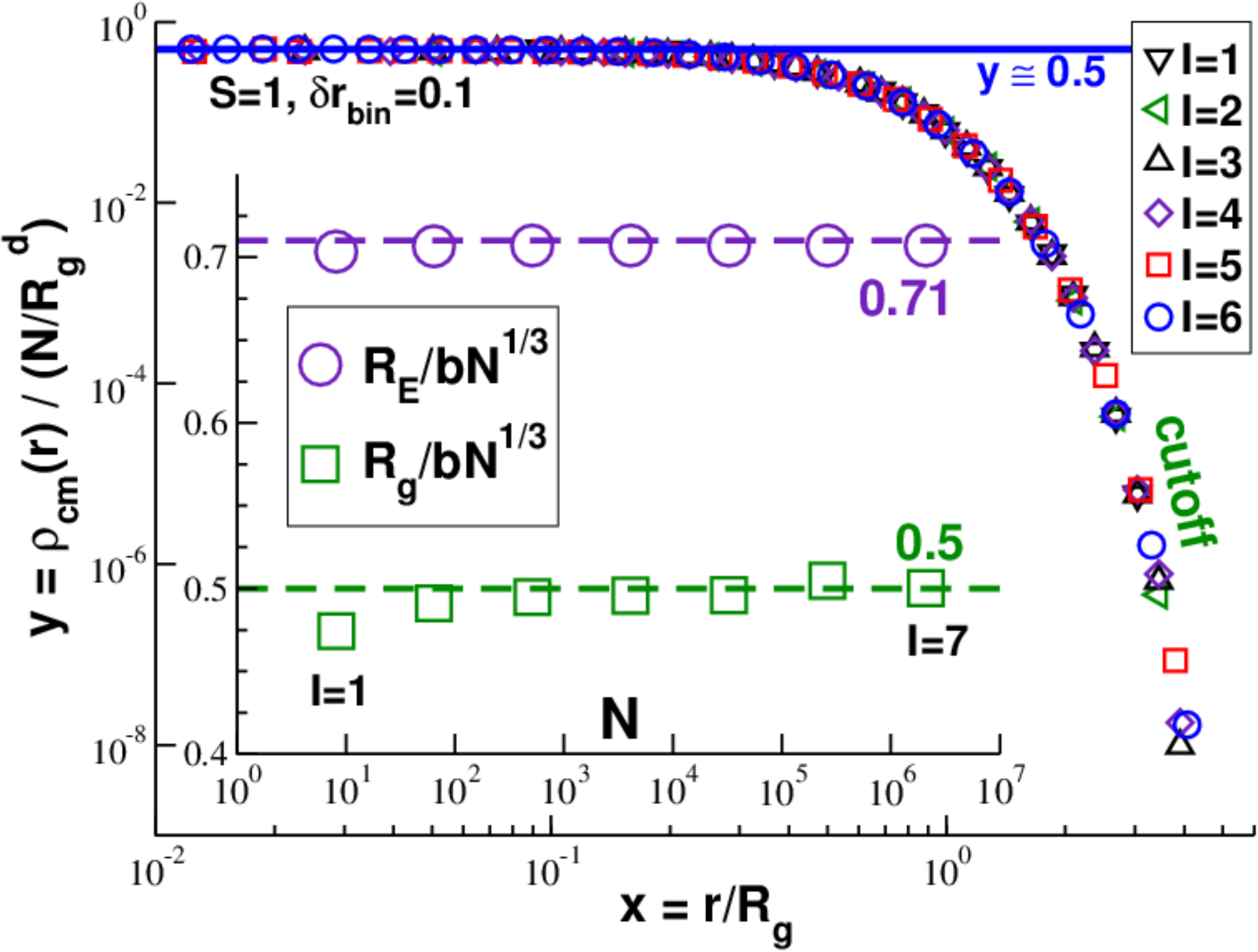}}}
%\vspace*{-1.5cm}
\caption{Inset:
Rescaled root-to-end distance $\Rend/bN^{1/3}$ and radius of gyration $\Rgyr/bN^{1/3}$ for different $N(I)$. 
Main panel:
Self-density $\rcm(r)$ around chain center of mass for different $I$.
The data collapse if rescaled as indicated using $N(I)$ and $\Rgyr(I)$.
}
\label{fig_compact}
\end{figure}

\subsection{Compactness}
\label{stat_compact}

\paragraph*{Typical chain size.}
We demonstrate first that our hyperbranched trees become indeed compact, i.e. eqn~(\ref{eq_df}) holds, 
and this rapidly after one or two iterations.
The data presented in Fig.~\ref{fig_compact} have been obtained by MC simulations.
The typical chain size $R$ of the trees shown in the inset has been characterized using the 
root-mean-squared distance $\Rend$, as sketched in Fig.~\ref{fig_sketch_detail},
and the standard radius of gyration $\Rgyr$ of the chain.\cite{RubinsteinBook}
The ratio $R/bN^{1/3}$ is seen to be perfectly constant for both observables.
Compared to the sluggish crossover of ring melts,
our simple model thus becomes rapidly compact.
We remind \cite{dendgauss} that the Gaussian chain statistics along all chain contours 
implies $\Rend^2=b^2 S G$ and 
\begin{equation}
\Rgyr^2 = \frac{1}{2} \sum_{s=0}^{\smax} w(s) \Rseg^2
\mbox{ with } \Rseg^2 = b^2 s
\label{eq_Rg_ws}
\end{equation}
and $w(s)$ being the histogram of curvilinear distances discussed in Section~\ref{stat_ws}.
Both lengths $\Rend$ and $\Rgyr$ can thus be obtained without explicit simulations.

\paragraph*{Center of mass self-density $\rcm(r)$.}
The self-density $\rcm(r)$ at a distance $r$ from the chain center of mass is presented 
in the main panel. Plotting the rescaled density $y=\rcm(r)/(N/\Rgyr^d)$ as a function of 
the dimensionless distance $x=r/\Rgyr$ allows to collapse the data for all iterations $I$. 
As already stated $N/\Rgyr^d \sim N^0$, i.e. the plateau $\rcm(r\approx 0) \approx 4$ 
for $x \ll 1$ does {\em not} at all 
depend on $N$.\footnote[6]{The deviation from this scaling observed for melts of rings\cite{MWC96,KK11a} 
is due to insufficient chain lengths and the sluggish crossover to the compact limit
in these systems.\cite{obukhovmodel}}
We note finally that the self-density decreases as 
$\rcm(r) \sim N/\Rgyr^d \sim 1/\sqrt{S}$ with the spacer length (Table~\ref{tab_S}).

\subsection{Distribution $w(s)$}
\label{stat_ws}
\paragraph*{Definitions.}
Central properties characterizing the monomer connectivity are the distributions
\begin{equation}
w_k(s) \equiv \frac{1}{N} \sum_{l=1}^N \delta(s - s_{kl}) \mbox{ and }
w(s) \equiv \frac{1}{N} \sum_{k=1}^N w_k(s) 
\label{eq_wsdef}
\end{equation}
for $0 \le s \le \smax$ with $s_{kl}$ denoting the arc-length between the monomers $k$ and $l$.
$N w_k(s)$ counts the number of beads with curvilinear distance $s$ from a specific monomer $k$
and $N w(s)$ corresponds to the mean number of beads averaged over all reference monomers $k$.
Both distributions are normalized, i.e. $\sum_{s=0}^{\smax} w_k(s) = \sum_{s=0}^{\smax} w(s) = 1$. 
For consistency with earlier work \cite{dendgauss} we have included here the reference monomer 
$k$ at $s=0$ into the distributions. This is accounted for by the normalization. Trivially, 
$w(s=0) = 1/N$. (For some properties considered below it is more useful to exclude the reference 
monomer and to only consider distances $s > 0$. This leads to a renormalization factor $1/(1-1/N)$ 
if moments with $s > 0$ are considered using the histogram $w(s)$ defined for $s \ge 0$.) 
The first moment 
\begin{equation}
\frac{1}{2} \sum_{s} s w(s) \equiv \Wwiener/N^2
\label{eq_Wwiener}
\end{equation}
of the distribution, listed in Table~\ref{tab_I} and Table~\ref{tab_S}, defines the Wiener index 
$\Wwiener$.\cite{Wiener47,nitta94} According to eqn~(\ref{eq_Rg_ws}) this implies 
$\Wwiener/N^2 = (\Rgyr/b)^2$ for Gaussian trees.
Associated to the distributions $w_k(s)$ and $w(s)$ are the sums 
\begin{equation}
n_k(s) \equiv N \sum_{s'=0}^s w_k(s') \mbox{ and }
n(s) \equiv \frac{1}{N} \sum_{k=1}^N n_k(s) 
\label{eq_ns_def}
\end{equation}
measuring, respectively, the total mass attached within an arc-length $s$ to a specific monomer $k$
(as sketched in Fig.~\ref{fig_sketch_detail}) and the corresponding $k$-average.
Due to the normalization of $w_k(s)$ and $w(s)$ we have $n_k(s=\smax) = n(s=\smax) = N.$
A power-law ansatz $w(s) \sim s^{\alpha-1}$ implies $n(s) \sim s^{\alpha}$. Since 
\begin{equation}
n(s) \approx \rho R(s)^{d} \approx S (s/S)^{d/2} 
\label{eq_alpha}
\end{equation}
must hold for a self-similar compact network with Gaussian chain statistics, 
this implies $\alpha=d/2$. 

\begin{figure}[t]
\centerline{\resizebox{0.9\columnwidth}{!}{\includegraphics*{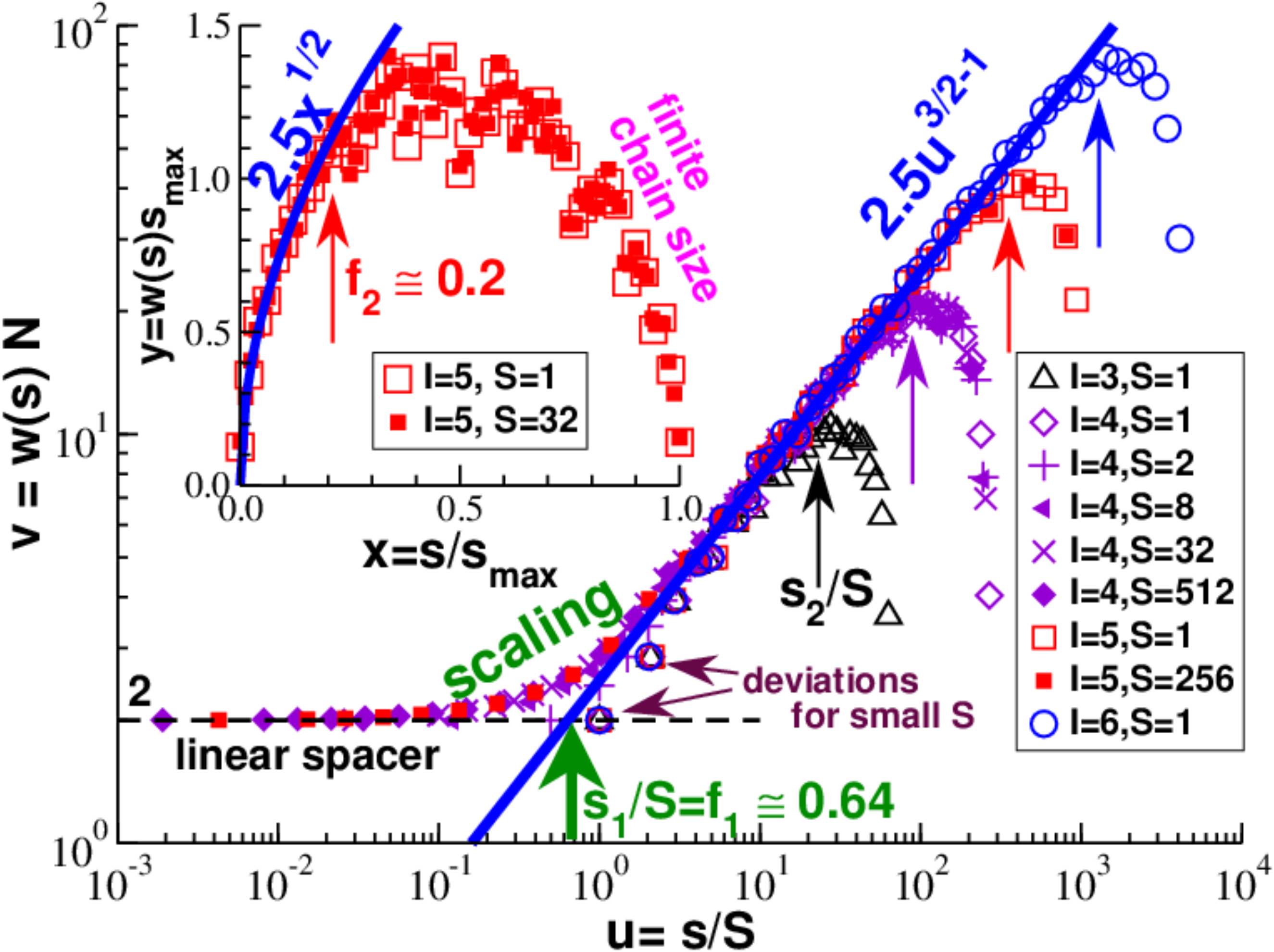}}}
\caption{Distribution $w(s)$ measuring the number of monomer pairs with curvilinear distance $s$.
Main panel:
Double-logarithmic representation of $v = w(s) N$ {\em vs.} $u=s/S$  
demonstrating eqn~(\ref{eq_ws_power}) for large $I$ (bold solid line).
The bold vertical arrow marks the crossover $\sone = S (2/\cws)^2$ between the linear spacer limit
(horizontal dashed line) and the intermediate power regime (bold solid line), all other vertical arrows
mark the breakdown of the power law at $\stwo=\ftwo \smax$.
Inset: 
Linear representation of $y = w(s) \smax$ {\em vs.} $x= s/\smax$ for $I=5$. 
}
\label{fig_ws}
\end{figure}

\paragraph*{Three regimes of $w(s)$.}
Various distributions $w(s)$ are presented in Fig.~\ref{fig_ws}. 
There are basically three regimes:
{\em (i)} the linear spacer limit with 
\begin{equation}
w(s) = 2/N \mbox{ for } s \ll \sone \equiv \fone S \mbox{ with } \fone \approx 0.64,
\label{eq_ws_spacer}
\end{equation}
{\em (ii)} 
the power-law regime expected according to eqn~(\ref{eq_alpha}) 
for marginal compact networks where 
\begin{equation}
w(s) = \frac{\cws}{N} (s/S)^{3/2-1} = \frac{\cws}{\smax} (s/\smax)^{3/2-1}
\label{eq_ws_power}
\end{equation}
holds (with eqn~(\ref{eq_Ntot}) being used in the second step) 
for $\sone \ll s \ll \stwo \equiv \ftwo \smax$ with $\cws \approx 5/2$ and 
$\ftwo \approx 0.2$ and
{\em (iii)} 
the final cutoff beyond $\stwo$ where non-universal finite chain size effects become dominant.
We comment now on the scaling of these regimes.

\paragraph*{Scaling with $s/S$.}
Using a double-logarithmic representation the main panel presents a broad range of $I$ and $S$ tracing 
$v=w(s) N$ as a function of $u=s/S$. This allows to collapse the data for small and intermediate $s$. 
Deviations from this scaling are seen for small $S < 8$ and $u \approx 1$ due to too small spacer 
lengths and for $s > \stwo$ as marked by thin vertical arrows.
The dashed horizontal line indicates the limit eqn~(\ref{eq_ws_spacer}) for long spacer chains
expressing the fact that each monomer has essentially {\em two} neighbors at curvilinear distance $s$.
As expected from the scaling argument eqn~(\ref{eq_alpha}), it is seen that eqn~(\ref{eq_ws_power}) 
holds over more than two decades for the largest $I$ presented (bold solid line).\footnote[7]{Interestingly,
essentially the {\em same} power-law exponent has been fitted in recent simulations of three-dimensional melts
of self-avoiding branched polymers.\cite{Everaers17,Everaers17b} This suggests that marginal compactness remains 
relevant if excluded volume is switched on provided that the connectivity is annealed.}
The bold vertical arrow at $\sone/S \equiv \fone = (2/\cws)^2 \approx 0.64$ marks the crossover of the first two regimes.
Using eqn~(\ref{eq_Rg_ws}) and eqn~(\ref{eq_Ntot}) one verifies that eqn~(\ref{eq_ws_power}) 
is consistent with $\Rgyr \approx b\sqrt{S} \ (N/S)^{1/d}$.

\paragraph*{Scaling with $s/\smax$.}
Using linear coordinates $y=w(s)\smax$ is plotted in the inset as a function of $x=s/\smax$. 
Data for $I=5$ and two spacer lengths $S$ are shown. 
The rescaling with $\smax$ allows quite generally to collapse data for large arc-lengths $s \gg S$. 
It is seen that $w(s)$ is a non-monotonous distribution, having a maximum at $x \approx 1/2$ 
and vanishing for $x\to 0$ and $x \to 1$. 
The vertical arrow marks the arc-length $\stwo/\smax = \ftwo \approx 0.2$ above which the power law 
eqn~(\ref{eq_ws_power}) becomes inaccurate. The latter value depends somewhat on the criterion used.

\begin{figure}[t]
\centerline{\resizebox{0.9\columnwidth}{!}{\includegraphics*{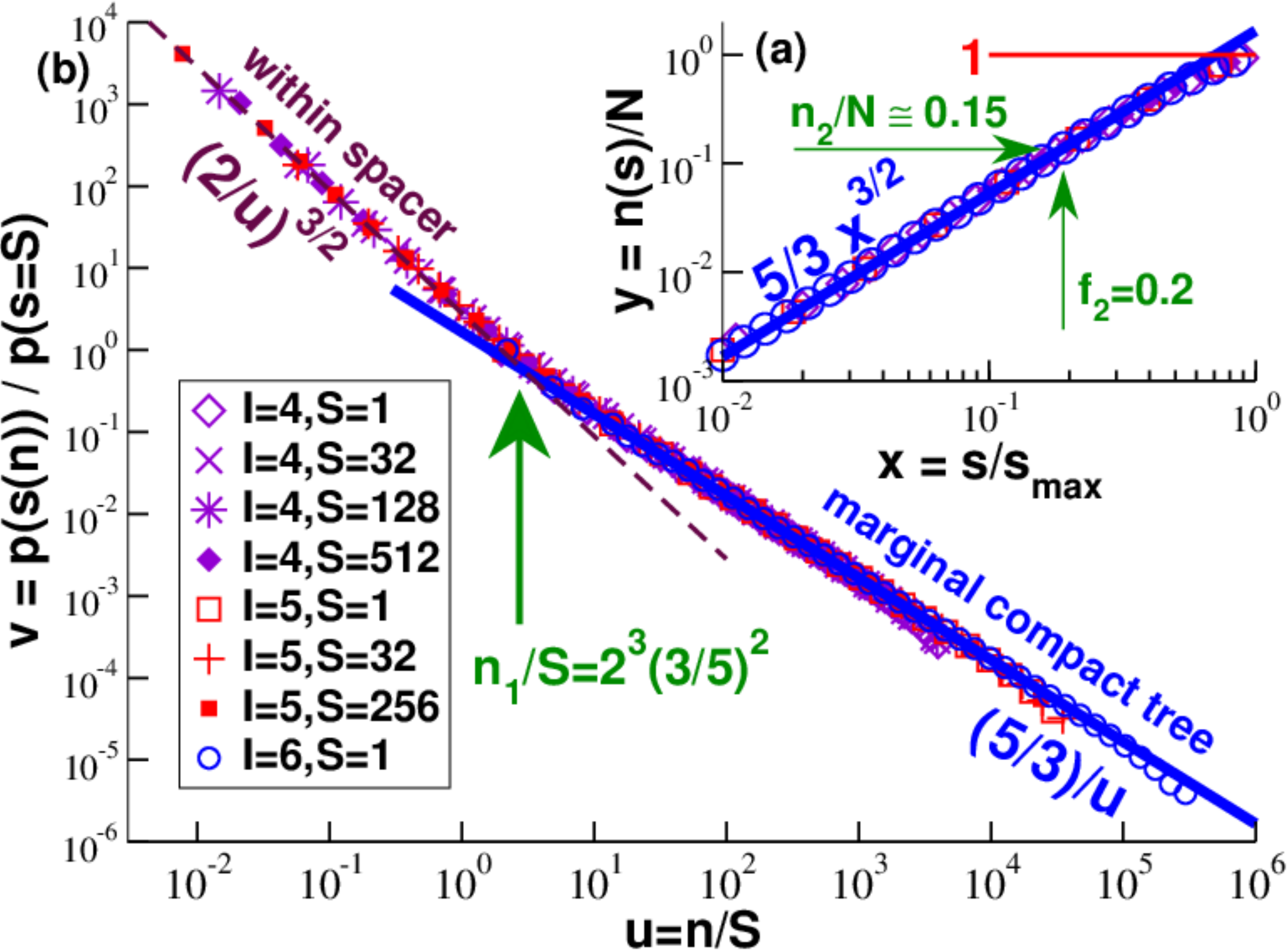}}}
\caption{Return probability $p(n)$ for $I=4$, $5$ and $6$. Open symbols indicate $S=1$.  
{\bf (a)} 
Rescaled mass $y=n(s)/N$ {\em vs.} $x=s/\smax$ with $n(s)$ being 
obtained using eqn~(\ref{eq_ns_def}). 
{\bf (b)}
Double-logarithmic representation of the rescaled return probability 
$v=p(n(s))/p(s=S)$ {\em vs.} $u=n/S$ allowing the scaling of all data for $s \ll \stwo$.
The linear chain power law $(2/u)^{3/2}$ is seen for large $S$ and small $u$ (dashed line).
The bold line indicates the power law $(5/3)/u$ for the self-similar marginal compact regime.
}
\label{fig_pn}
\end{figure}

\subsection{Return probability}
\label{stat_pn}

We turn now to the return probability $p(n)$ mentioned in the Introduction.
Let us consider two monomers $k$ and $l$ being an arc-length $s$ apart
as shown in Fig.~\ref{fig_sketch_detail} for $s=4$.
Using the distribution $P(\rvec,s)$ given by eqn~(\ref{eq_Grsgauss})
the return probability is simply 
\begin{equation}
p(s) \equiv P(\rvec=\underline{0},s) =  \left(d/2\pi b^2 s\right)^{d/2} \mbox{ for all }  s.
\label{eq_preturn_s}
\end{equation}
It is now customary and useful (see Section~\ref{stat_contact}) 
to express the return probability in terms of the typical mass $n(s)$.
According to eqn~(\ref{eq_ns_def}), $n(s)$ must be a monotonously increasing function and 
its inverse $s(n)$ may thus be obtained from the distribution $w(s)$.
This is confirmed in panel (a) of Fig.~\ref{fig_pn} where $y=n/N$ is plotted as a function of $x=s/\smax$.  
Following eqn~(\ref{eq_ws_power}) we have
\begin{equation}
n(s)= (2\cws/3) \ S (s/S)^{3/2} \mbox{ for } \sone \ll s \ll \stwo.
\label{eq_nsint}
\end{equation}
This corresponds to $y = (5/3)x^{3/2}$ as indicated by the bold solid line. 
A continuous crossover to unity is observed for larger $x$ (finite chain size).
Panel (b) of Fig.~\ref{fig_pn} presents the distribution $p(s(n))$ obtained using eqn~(\ref{eq_preturn_s}) 
and $s(n)$. The data are brought to collapse by plotting $v = p(s(n))/p(s=S)$ as a function
of $u=n/S$, i.e. we take the return probability of the spacer chain as reference.
Due to eqn~(\ref{eq_ws_spacer}) we have $n(s)=2s$ which implies $v=(2/u)^{3/2}$
in the small-$u$ limit (dashed line). 
Using again eqn~(\ref{eq_nsint}) this leads to $v= (5/3)/u$ for large $u$ (bold line). 
The return probability thus decreases as $p(n) \sim 1/(\sqrt{S}n)$ within the $n$-range
\begin{equation}
\none = 2\sone = 2 \fone \ S \ll n \ll \ntwo = (2\cws/3) \ftwo^{3/2} \ N
\label{eq_nrange}
\end{equation}
where we have used eqn~(\ref{eq_Ntot}) for the upper limit. 
The limits $\none$ and $\ntwo$ are indicated by arrows in Fig.~\ref{fig_pn}.
Note that the representation of panel (b) masks somewhat that the scaling fails above $\ntwo$.
We are now ready to embark on the key issue of this paper.

\subsection{Self-contact density}
\label{stat_contact}

\begin{figure}
\centerline{\resizebox{0.9\columnwidth}{!}{\includegraphics*{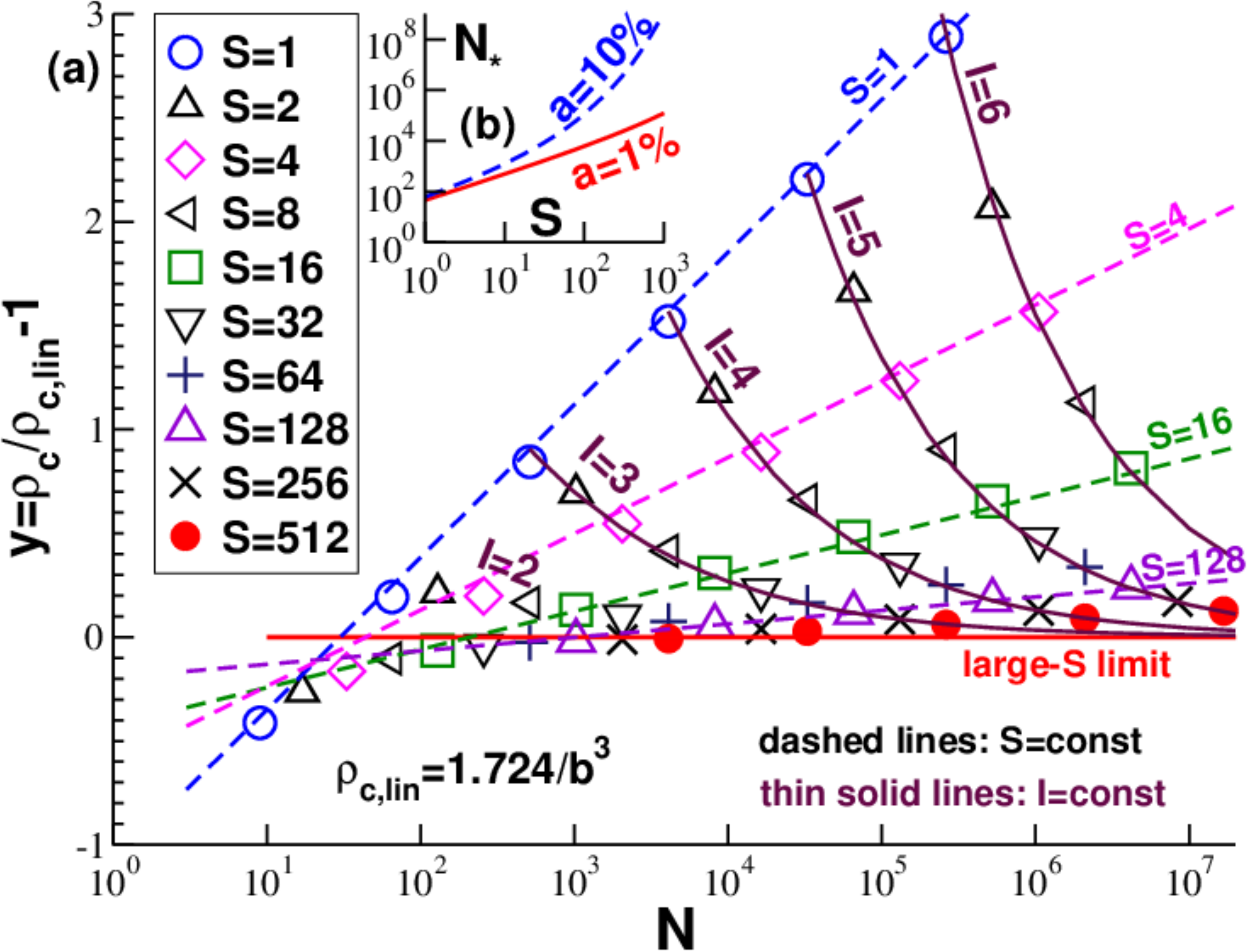}}}
\caption{Self-contact density $\rcon$:
{\bf (a)} $y=\rcon/\rconlin-1$ {\em vs.} $N$ for a broad range of spacer lengths $S$
using $\rconlin=1.724/b^3$ as reference. In agreement with eqn~(\ref{eq_rconfit}) the data diverge
logarithmically with mass $N$ at constant spacer length $S$ (thin dashed lines) and decays as 
$y \sim 1/\sqrt{S}$ with spacer length $S$ at constant iteration $I$ (thin solid lines). 
{\bf (b)} Upper limit $\Nstar(S)$ according to eqn~(\ref{eq_Nupper}) for $a=10\%$ and $a=1\%$.
}
\label{fig_contact}
\end{figure}

\paragraph*{Observable.}
The self-contact density $\rcon$ presented in Fig.~\ref{fig_contact} has been obtained using the sum
\begin{equation}
\rcon = \frac{1}{1-1/N} \sum_{s=1}^{\smax} N w(s) \times p(s)
\label{eq_ws2rcon}
\end{equation}
taking advantage of the distribution $w(s)$ characterizing the tree connectivity.
The prefactor $1/(1-1/N)$ appears for normalization reasons (Section~\ref{stat_ws}) since the sum 
runs from $s=1$ excluding the reference monomer at $s=0$. This factor becomes rapidly irrelevant 
and will be ignored below. Since according to eqn~(\ref{eq_ns_def}) $N w(s) = \ddiff n(s)/\ddiff s$,
the above sum (integral) over $s$ becomes equivalent in the continuum limit to the integral over $n$
\begin{equation}
\rcon \approx \int_{n=1}^N \ddiff n \ p(n) \approx \sum_{n=1}^N p(n), 
\label{eq_Pnsum}
\end{equation}
i.e. the weight $w(s)$ needed in eqn~(\ref{eq_ws2rcon}) naturally drops out if $n$ is used as variable. 
We emphasize that being applicable for arbitrary tree networks, eqn~(\ref{eq_Pnsum}) is more general 
than the (formally identical) well-known expression for unbranched (linear or closed loop) polymer 
chains.\cite{KK11a} 
The self-contact density $\rcon$ is thus equivalent to the surface below the data in Fig.~\ref{fig_pn}.

\paragraph*{Scaling of data.}
The main panel (a) of Fig.~\ref{fig_contact} presents the rescaled self-contact density 
$y=\rcon/\rconlin-1$ as a function of $N$ for a broad range of spacer lengths $S$. 
The self-contact density $\rconlin \approx 1.724/b^3$ for arbitrarily long linear Gaussian chains
is used as reference. See Appendix~\ref{app_contact} for details. 
As shown by the dashed lines for a constant spacer length $S$ and by the thin solid lines for
a constant iteration $I$ our data are well described using 
\begin{equation}
y \equiv \frac{\rcon}{\rconlin} -1 = \frac{\cone}{\sqrt{S}}\left[\log(N/S)-\ctwo\right]
\label{eq_rconfit}
\end{equation}
with $\cone \approx 0.319$ and $\ctwo \approx 3.4$. 
As shown in Appendix~\ref{app_contact}, the first coefficient $\cone$ is known exactly.
It stems from the contribution of the intermediate marginal compact regime of $w(s)$
which must dominate the integral eqn~(\ref{eq_ws2rcon}) for asymptotically large trees.
For large $I$ and constant $S$ the self-contact density thus diverges logarithmically as 
$\rcon/\rconlin \approx \cone \log(N)/\sqrt{S}$ with $N$, while for constant $I$ and 
increasing $S$ the data approach rapidly the linear chain reference (bold horizontal solid line).
The coefficient $\ctwo$, summarizing all subdominant corrections to the asymptotic behavior, 
is more difficult to predict due to the various crossovers and has been fitted. 
More details on this minor technical point can be found at the end of Appendix~\ref{app_contact}.

\paragraph*{Upper mass bound.}
The central consequence of eqn~(\ref{eq_rconfit}) is that if a local relative density
fluctuation $y = a$ is physically just acceptable, an allowed chain mass $N$ must satisfy 
the inequality
\begin{equation} 
N \ll \Nstar(S) \equiv S \exp\left( \frac{\sqrt{S} a}{\cone} + \ctwo \right).
\label{eq_Nupper}
\end{equation}
As may be seen from panel (b) of Fig.~\ref{fig_contact} for $a = 10\%$ and $a=1\%$, 
the upper limit $\Nstar$ increases dramatically with $S$, i.e. the marginal compact model 
provides a physically acceptable model over several orders of magnitude.
Please note that the indicated parameters $a$ are extremely conservative considering 
that in normal excluded volume polymer fluids the local density around a reference monomer may 
fluctuate by a factor of order unity.

\subsection{Radial intrachain pair density distribution}
\label{stat_pair}

\begin{figure}[t]
\centerline{\resizebox{0.9\columnwidth}{!}{\includegraphics*{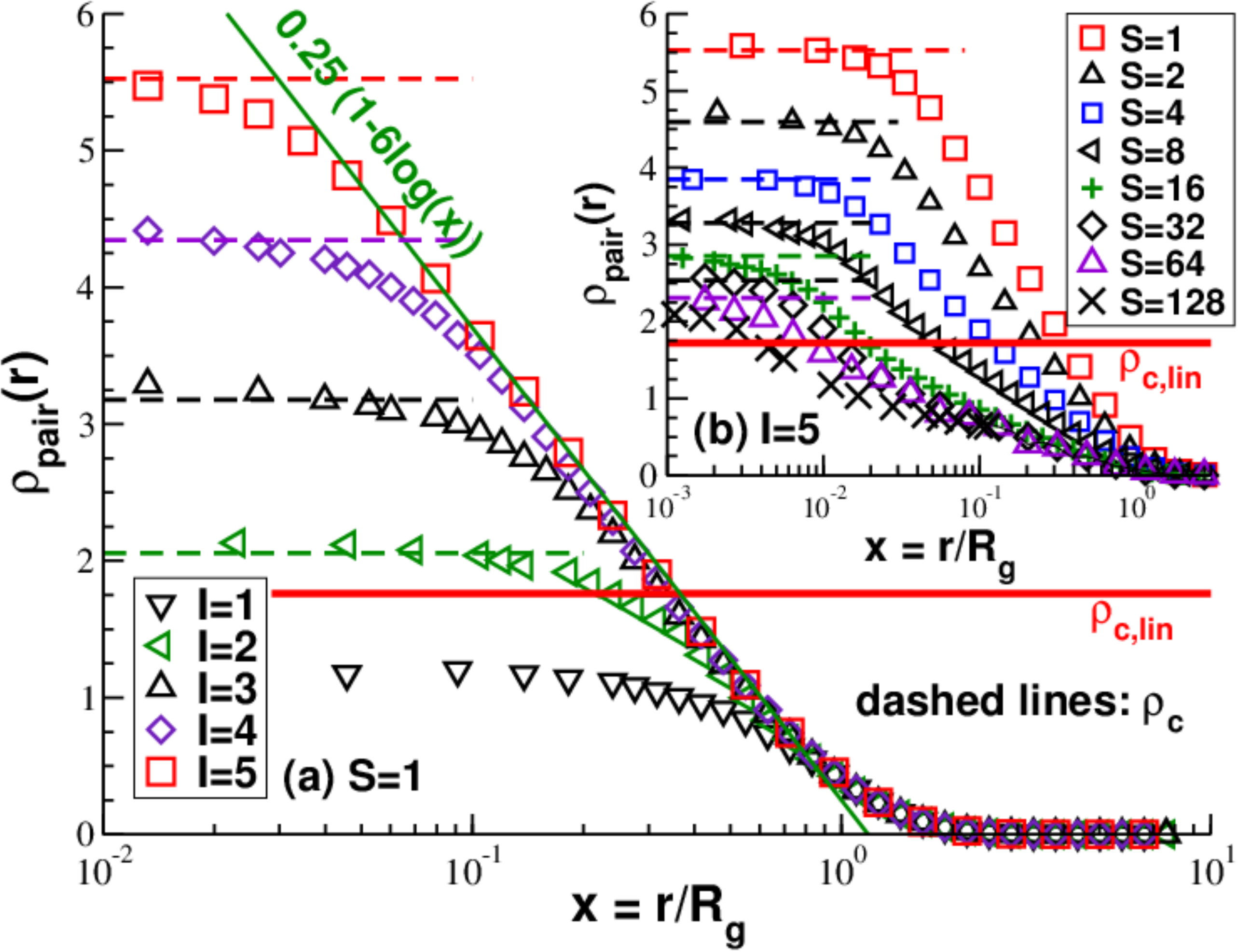}}}
\caption{Radial intrachain pair density distribution $\rpair(r)$:
{\bf (a)} different $I$ at constant spacer length $S=1$,
{\bf (b)} different $S$ at constant $I=5$.
The bold horizontal lines indicate the self-contact density $\rconlin$ for long linear chains. 
As shown by the dashed horizontal lines $\rcon \approx \rpair(r\approx 0)$.
}
\label{fig_pair}
\end{figure}

The self-contact density $\rcon$ has been determined in Section~\ref{stat_contact} using the histogram $w(s)$ 
without explicit computer simulation. This has allowed us to scan over a huge range of $N$.
In addition to this we have computed $\rcon$ from the $r \to 0$ limit of the directly simulated
radial pair density distribution $\rpair(r)$. We remind that $\rpair(r)$ may be obtained by radially 
averaging over
\begin{equation}
\rpair(r) = \frac{1}{N}\sum_{k=1}^N \sum_{l\ne k} \langle \delta(\rvec-(\rvec_l-\rvec_k))\rangle.
\label{eq_rpair_def}
\end{equation}
The data presented in Fig.~\ref{fig_pair} have been computed by means of MC simulations 
using pivot moves.\cite{dendgauss}
As shown by the dashed horizontal lines, $\rcon \approx \rpair(r \approx 0)$ holds as expected.
We use a half-logarithmic representation with a rescaled horizontal axis $x=r/\Rgyr$. 
As shown in the main panel (a) for a spacer length $S=1$, $\rpair(x)$ differs quite strongly for small $S$ 
from the average density $\rcm(x)$ around the chain center of mass considered in Fig.~\ref{fig_compact}:
The data do not scale and increase logarithmically with decreasing $x$.
Note that $\rpair(x)$ increases strongly above the self-contact density $\rconlin$ for 
long linear Gaussian chains indicated by the solid horizontal line.
Spacer length effects are considered in the inset of Fig.~\ref{fig_pair} for $I=5$.
It is seen that $\rpair(x)$ decreases with $S$ approaching from above an $S$-independent asymptote. 
We shall further analyse these pair correlations in the next subsection where we discuss the related, 
but experimentally more relevant intramolecular structure factor $F(q)$.

\begin{figure}[t]
\centerline{\resizebox{0.9\columnwidth}{!}{\includegraphics*{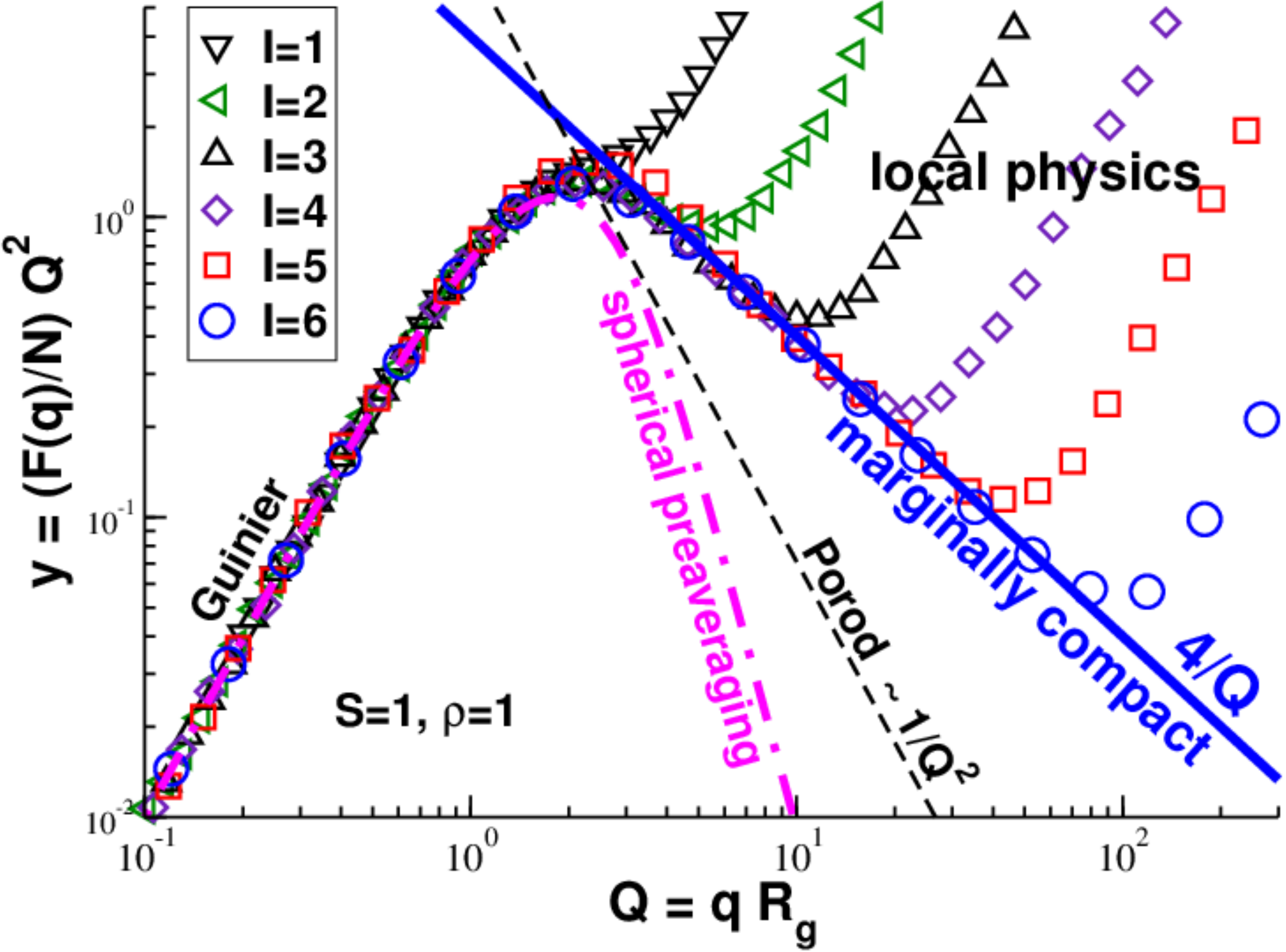}}}
\caption{Kratky representation $y=(F(q)/N) Q^2$ {\em vs.} $Q=q\Rgyr$ of the static 
intrachain form factor $F(q)$ for different iterations $I$ as indicated in the figure 
with $q$ being the wavevector and $\Rgyr(N)$ the radius of gyration. 
The trees become {\em marginally} compact, i.e. $y \sim 1/Q$ for sufficiently large $I$ 
(bold solid line). The dashed line indicates the standard Porod scaling for a compact object 
with a well-defined non-fractal surface.
The dash-dotted line has been obtained according to eqn~(\ref{eq_Fq_preaver})
of Appendix~\ref{app_Fq}.
}
\label{fig_Fq}
\end{figure}

\subsection{Static structure factor}
\label{stat_Fq}

Conformational properties of polymer chains can be determined experimentally by means of light, 
small angle X-ray or neutron scattering experiments.\cite{BenoitBook,RubinsteinBook}
This allows to extract the coherent intramolecular structure (form) factor 
$F(q) = \frac{1}{N} \langle || \sum_{k=1}^{N} \exp\left(\text{i} \qvec \cdot \rvec_k \right) ||^2 \rangle$
with $\rvec_k$ the monomer position and $\qvec$ the wavevector. 
For polymer networks with Gaussian chain statistics the form factor may be computed directly
using \cite{dendgauss}
\begin{equation}
F(q) = \sum_{s=0}^{\smax} w(s) P(q,s)
\label{eq_Fq_ws}
\end{equation}
with $P(q,s)$ being the Fourier transform of the segment size distribution $P(\rvec,s)$.
Since for Gaussian chains $P(q,s) = \exp(-(a q)^2 s)$ with $a \equiv b/\sqrt{2d}$, 
the form factor is readily computed yielding, as one expects, 
within numerical accuracy exactly the same results as obtained using the configuration ensembles
computed by means of MD and MC simulations.
Data for $S=1$ are presented in Fig.~\ref{fig_Fq} where a Kratky representation is used.\cite{BenoitBook}
We remind that for large $N$ and small $q \equiv ||\qvec||$ the radius of gyration $\Rgyr$, 
as one measure of the tree size, must become the only relevant length scale.
The form factor thus scales as $F(q)/N = f(Q)$ with $Q = q\Rgyr$ being the reduced wavevector and 
$f(Q)$ a universal scaling function with $f(Q) = 1 - Q^2/d$ in the ``Guinier regime" for $Q \ll 1$
as seen on the left side of the figure.
The opposite large-$q$ limit probes the density fluctuations within the spacer chains and on the scale
of the monomers. The data thus do not scale with $Q$ just as the pair correlation density $\rpair(r)$
in Fig.~\ref{fig_pair} did not scale for small $r/\Rgyr$. 
Details on this limit are given in Appendix~\ref{app_Fq}.
While one expects $F(q)/N \sim 1/Q^{\df}$ in the intermediate wavevector regime of ``open" fractal objects
($\df < d$),\cite{DoiEdwardsBook}
compact fractals are described by the ``generalized Porod law"\cite{BenoitBook,MSZ11}
\begin{equation}
F(q)/N \sim 1/Q^{2d-\ds} \ \mbox{ for } \df=d, \ds < d.
\label{eq_GPL}
\end{equation}
By matching these well-known limits, this demonstrates the key relation eqn~(\ref{eq_Fq_margcomp}) 
for marginally compact objects. Alternatively, eqn~(\ref{eq_Fq_margcomp}) can be confirmed
using eqn~(\ref{eq_Fq_ws}) and eqn~(\ref{eq_ws_power}).
As seen in Fig.~\ref{fig_Fq}, the data approach as expected with increasing $I$ the slope 
$y \approx 4/Q$ (bold line). This confirms the claimed {\em marginal} compactness of our trees. 
The standard Porod scattering \cite{BenoitBook} of compact objects with a smooth surface ($\ds=2$) 
corresponds instead to a much steeper power-law envelope (dashed line).

\section{Dynamical properties}
\label{sec_dyna}
\begin{figure}[t]
\centerline{\resizebox{0.9\columnwidth}{!}{\includegraphics*{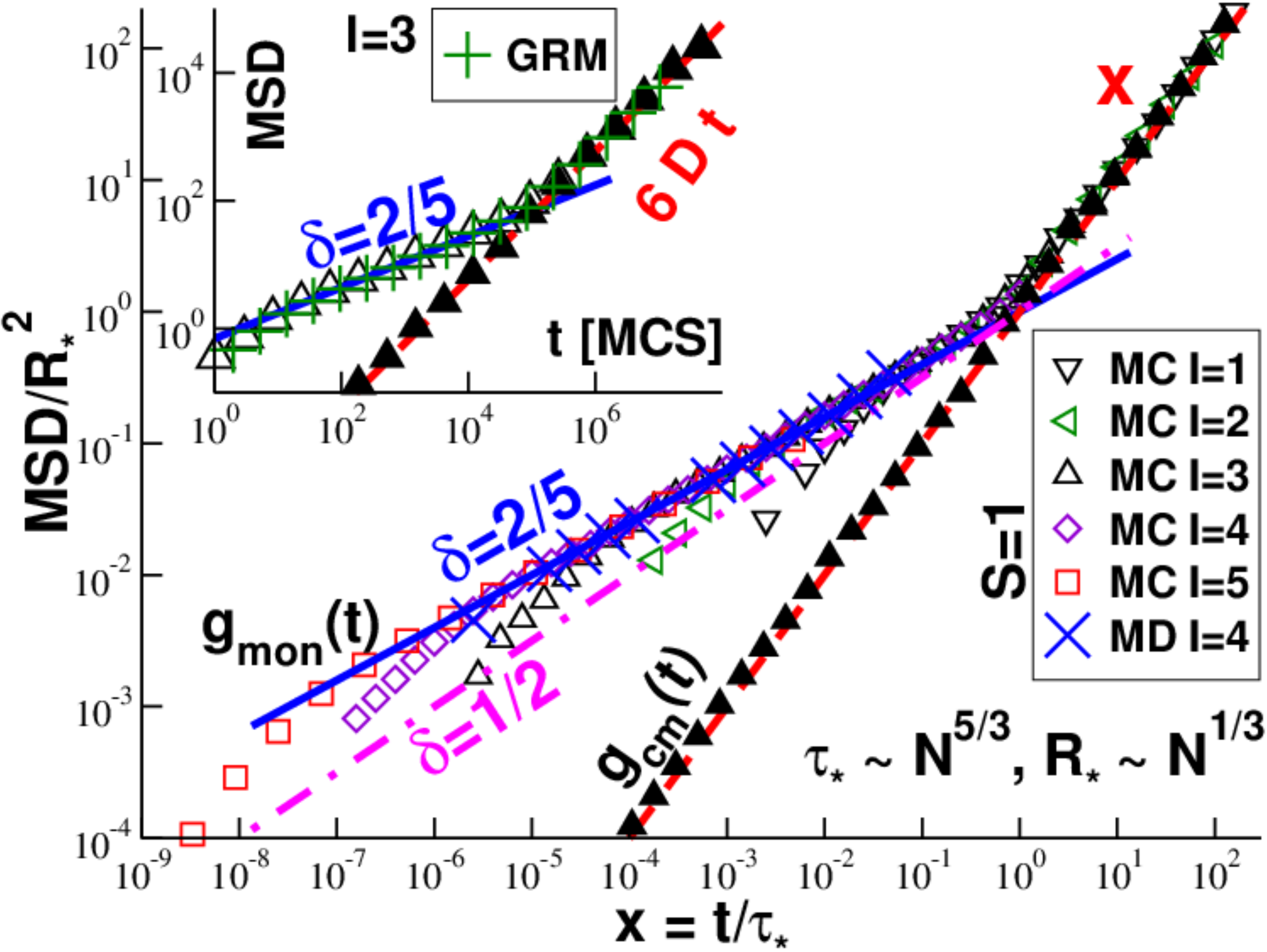}}}
\caption{Scaling of mean-square displacements (MSD).
Filled triangles indicate the center-of-mass MSD $\gcms$ for $I=3$, open symbols the monomer MSD $\gmon$,
the bold solid lines the exponent $\delta=2/5$ from eqn~(\ref{eq_dyna_gmon}) with $\nu=1/3$.
Inset: $\gmon$ and $\gcms$ {\em vs.} time $t$ in MC steps for $I=3$
computed using local MC jumps with $\delta r = 0.5$. Also indicated is the GRM prediction for $\gmon$ (pluses).
Main panel: Scaling collapse for $\gmon/\Rstar^2$ {\em vs.} $x = t/\Tstar$ for MC with different $I$
as indicated (open symbols) and MD data for $I=4$ (crosses).
\label{fig_MSD}
}
\end{figure}

\subsection{Introduction}
\label{dyna_intro}
We describe now some dynamical properties of our marginally compact trees focusing 
on a spacer length $S=1$. We remind that all excluded-volume, topological or hydrodynamic 
interactions between the $m=100$ chains of the simulation box at $\rho=1$ are switched off. 
The Gaussian chain connectivity is the only remaining potential.
For clarity of the presentation we focus on numerical data obtained using local MC jumps. 
Occasionally, we include the corresponding results obtained by means of MD simulations or 
using the generalized Rouse model (GRM). 
The principal goal is to show that the dynamics is of a generalized Rouse-type
characterized by an inverse fractal dimension $\nu \equiv 1/\df = 1/d$.

\subsection{Mean-square displacements}
\label{dyna_MSD}

Since the effective random forces acting on the monomers are uncorrelated,
a pure Fickian diffusion is expected for the center-of-mass of the chains.\cite{DoiEdwardsBook}
The corresponding mean-square displacement (MSD) $\gcms$ should thus scale as 
\begin{equation}
\gcms = 2d D t \ \mbox{ with } D = \kBT / \zeta N
\label{eq_dyna_gcm}
\end{equation}
being the diffusion coefficient and $\zeta$ the friction coefficient. 
As shown in Fig.~\ref{fig_MSD}, this is consistent with our simulations. 
We find that $\zeta \approx 44$ for MC and $\zeta = 10$ for MD using appropriate units.
(The latter value for the MD simulations is imposed by the friction coefficient of the Langevin thermostat.)
Having thus determined the central parameter $\zeta$, this allows us to compare the predictions of the 
GRM with our MC and MD simulations.
Since $R \sim N^{\nu}$, one expects a characteristic chain relaxation time \cite{DoiEdwardsBook}
\begin{equation}
\tauN \sim R^2/D \sim N^{\beta} \mbox{ with } \beta \equiv 1+2\nu
\label{eq_tauN}
\end{equation}
in agreement with eqn~(\ref{eq_app_d_s_d_f}).
The same scaling argument holds for the relaxation time $\taun \sim n^{\beta}$ of a 
subchain of mass $n$ around an arbitrary tagged monomer $k$ (Fig.~\ref{fig_sketch_detail}).
Compact (sub)chains ($\nu=1/d$) thus relax {\em faster} in $d > 2$ than Gaussian (sub)chains ($\nu=1/2$) of same mass. 

The scaling of the relaxation time may be verified using the monomer MSD $\gmon$ 
as shown in Fig.~\ref{fig_MSD} (open symbols).
For large times $t \gg \tauN$ the monomers must follow the chain center-of-mass, i.e. $\gmon$ is given by eqn~(\ref{eq_dyna_gcm}). 
A free diffusion $\gmon \sim t N^0$ is also observed for our MC simulations 
in the opposite limit of very small times $t \ll \taumon \sim N^0$.
(Depending on the friction constant $\zeta$ a ballistic regime with $\gmon \sim N^0t^2$ appears for 
our MD simulations in this time regime.)
The chain connectivity matters, however, in the intermediate time window for $\taumon \ll t \ll \tauN$. 
The anomalous diffusion in this regime can be understood using the standard scaling argument
\cite{DoiEdwardsBook}
\begin{equation}
\gmon \approx R^2 (t/\tauN)^{\delta} \sim N^0 \ \mbox{ for } \taumon \ll t \ll \tauN
\label{eq_dyna_gmon}
\end{equation}
with $\delta = 2\nu/\beta = 1/(1+1/2\nu)$.
Alternatively, one may obtain eqn~(\ref{eq_dyna_gmon}) from the 
center-of-mass mean-square displacement
\begin{equation}
\gcmsn \sim \frac{1}{n(t)} \times t \sim n(t)^{2\nu} \sim t^{\delta}
\end{equation}
of the $n(t) \sim t^{1/\beta}$ monomers dragged along by a reference monomer $k$ within a time window $t$.
As seen in Fig.~\ref{fig_MSD} the power law $t^{2/5}$ (bold solid lines) corresponding to 
$\nu=1/d$ perfectly fits the data.
As shown in the main panel of Fig.~\ref{fig_MSD}, all MC and MD data can be brought to collapse 
($t \gg \taumon$) using as dimensionless coordinates the rescaled time $x=t/\Tstar$ and 
the rescaled MSD $y=MSD/\Rstar^2$. The crossover size $\Rstar = 2.2 \Rgyr \sim N^{\nu}$ and 
the crossover time $\Tstar = \Rstar^2/6D \sim N^{\beta}$ have been chosen to match both asymptotic slopes at $x=1$. 

\subsection{Relaxation time of generalized Rouse modes}
\label{dyna_taup}

\paragraph*{Monomer MSD revisited.}
Due to the bilinearity of our model potential, eqn~(\ref{eq_Vspring}), 
the monomer MSD $\gmon$ presented in Fig.~\ref{fig_MSD} can be directly predicted 
using the GRM outlined in Appendix~\ref{app_theo}. 
As shown elsewhere,\cite{DoiEdwardsBook,schiessel98,gurtovenko05}
the MSD can be written as
\begin{equation}
\gmon = 2d D t \left( 1 + \sum_{p=1}^{N-1} \frac{1- \exp(-t/\taup)}{t/\taup}  \right) 
\label{eq_gmon_GRM}
\end{equation}
with $\taup$ being the relaxation time of the mode $p>0$. 
As shown in the inset of Fig.~\ref{fig_MSD}, eqn~(\ref{eq_gmon_GRM}) yields 
the same results (crosses) as the MC simulations (open triangles).
The relaxation times $\taup$ have been determined using $\taup = \taumon/\lamp$ 
with $\taumon = \zeta/K$ being a convenient constant and $\lamp$ the eigenvalues 
of the connectivity matrix $\mathbf{A}$.
While the local time scale $\taumon \sim N^0$ depends due to $\zeta$ on the simulation method, 
the same eigenvalues $\lamp$ characterize MC and MD simulations.

\begin{figure}[t]
\centerline{\resizebox{0.9\columnwidth}{!}{\includegraphics*{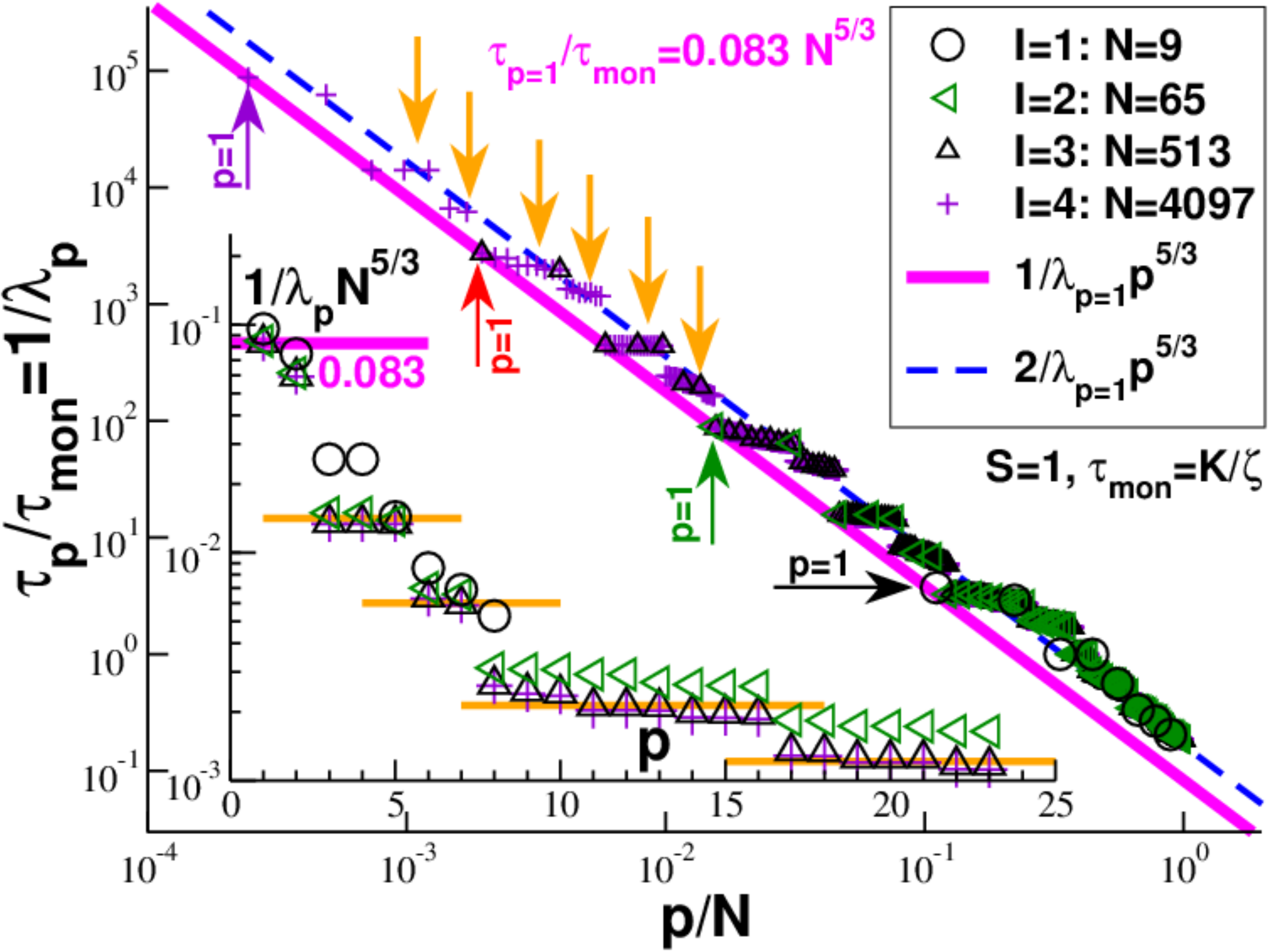}}}
\caption{Generalized Rouse relaxation times $\taup/\taumon=1/\lamp$ for four chain lengths $N(I)$.
The smallest eigenmode $p=1$ scales as $\tau_{p=1}/\taumon \approx 0.083 N^{5/3}$ (bold solid lines).
Inset: Half-logarithmic representation of $1/\lamp N^{5/3}$ {\em vs.} $p$ for the first 23 eigenmodes.
Degeneracies are marked by horizontal lines.
Main part: Double-logarithmic representation of $1/\lamp$ {\em vs.} $p/N$. 
}
\label{fig_taup}
\end{figure}

\paragraph*{Scaling of relaxation times $\taup$.}
The reduced relaxation times $\taup/\taumon$, i.e. the inverse eigenvalues $\lamp$, 
are shown in Fig.~\ref{fig_taup} for four different $I$.
We emphasize first that, as shown by the bold solid lines, the relaxation time of the
eigenmode $p=1$ increases as $\tau_{p=1}/\taumon \approx 0.083 N^{5/3}$ 
as one expects from eqn~(\ref{eq_tauN}) and $\tauN \approx \tau_{p=1}$. 
(Already the value for the tiny trees with $I=1$ comes close to this limit.)
The inset presents the first 23 eigenmodes using a half-logarithmic representation where 
$\taup/\taumax N^{5/3}$ is plotted as a function of $p$. All relaxation times $\taup$ are given in the 
main panel using a double-logarithmic representation with the rescaled mode number $p/N$ as horizontal axis.
The coordinates have been chosen to verify the expected data collapse with respect to the chain length $N$ 
in the regime each panel focuses on.
As seen from the inset, the relaxation times $\taup$ for small $p$ do not decrease continuously 
but with clearly separated steps which become more marked with increasing $N$. 
(The data become $N$-independent in this limit.) 
These degeneracies reflect the symmetries of the tree generated by the generator (Fig.~\ref{fig_sketch}).
Despite these degeneracies the relaxation times $\taup$ collapse if plotted {\em vs.} $p/N$
as shown in the main panel. While this collapse is only approximative for small $p/N$, 
as emphasized by the arrows pointing downwards, it becomes clearly better with increasing $N$ and $p/N$.
Importantly, the relaxation times for all $p$ are within a narrow band between $1/\lambda_{p=1}p^{5/3}$ 
(bold solid line) and $2/\lambda_{p=1}p^{5/3}$ (dashed line). It is seen that 
the relaxation times approach the upper bound with increasing $p/N$. 
One may thus describe the relaxation times by an effective power law
\begin{equation}
\taup/\taumon \approx c (N/p)^{\beta} \mbox{ with } \beta=1+ 2\nu =5/3
\label{eq_taup}
\end{equation}
as before and $c$ a numerical constant of order unity.
Importantly, if we replace the directly measured $\taup$ by eqn~(\ref{eq_taup})
with $c \approx 2$, one obtains the same MSD $\gmon$ as before (not shown).
We note finally that using the power law $\taup \sim (N/p)^{\beta}$ in eqn~(\ref{eq_gmon_GRM}) for times 
$t \ll \tauN \approx \tau_{p=1}$ one confirms by replacing the sum by an integral that $\gmon \sim t^{2\nu/\beta}=t^{2/5}$ 
in agreement with eqn~(\ref{eq_dyna_gmon}).

\begin{figure}[t]
\centerline{\resizebox{0.9\columnwidth}{!}{\includegraphics*{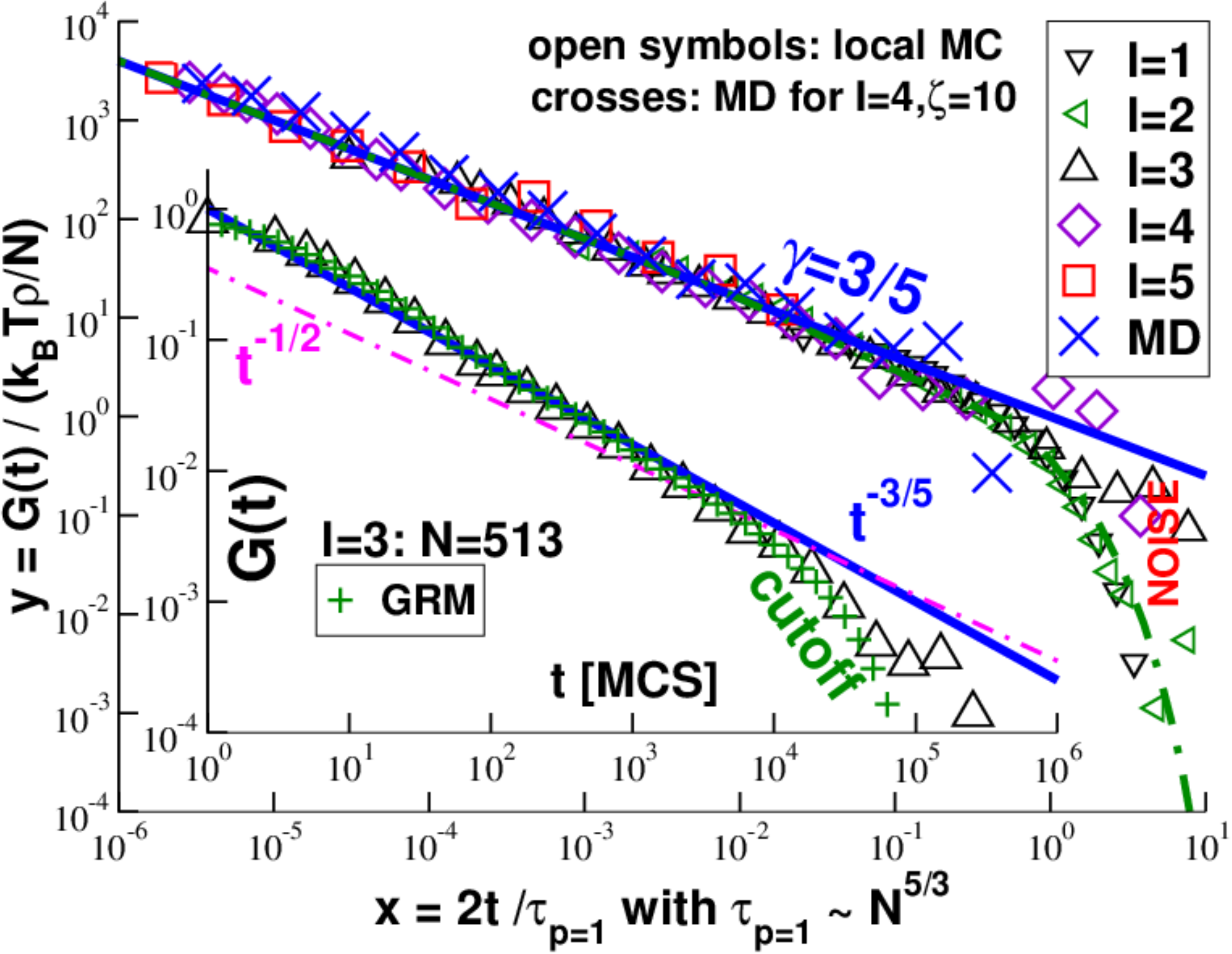}}}
\caption{Shear-stress relaxation modulus $G(t)$.
Inset: $G(t)$ for $I=3$ obtained with local jump MC and GRM prediction using eqn~(\ref{eq_Gt_GRM}). 
Main panel: $y= G(t) / (\kBT \rho/N)$ {\em vs.} reduced time $x = 2 t/\tau_{p=1}$. 
The predicted exponent $\gamma=3/5$ (bold lines) can be seen over several decades.
The dash-dotted line represents the GRM prediction $y \approx x^{-3/5}\exp(-x)$ for large $I$.
The cutoff for $x\gg 1$ is visible for the smaller trees.
}
\label{fig_Gt}
\end{figure}

\subsection{Shear-stress relaxation modulus}
\label{dyna_Gt}

A central rheological property characterizing the linear shear-stress response in fluids 
as well as in solids is the shear relaxation modulus $G(t) \equiv \delta \sigma(t)/\delta \gamma$ 
defined as the ratio of observed shear-stress increment $\delta \sigma(t)$ and applied shear-strain 
$\delta \gamma$.\cite{RubinsteinBook} 
Due to the simplicity of our Hamiltonian one may directly compute $G(t)$ using the 
(slightly rewritten) relation eqn~(4.158) from ref.~\cite{DoiEdwardsBook} 
\begin{equation}
G(t) = \kBT \nsp/V \ \sum_{p=1}^{N-1} \exp(-2 t/\taup)
\label{eq_Gt_GRM}
\end{equation}
with $\nsp$ being the number of springs and $\taup$ the mode relaxation time described above,
Fig.~(\ref{fig_taup}).
An example for $I=3$ is given in the inset of Fig.~\ref{fig_Gt} (pluses).
Using the shear-stress autocorrelation function\cite{AllenTildesleyBook} we have additionally 
obtained $G(t)$ from our equilibrium MC and MD simulations as also shown in Fig.~\ref{fig_Gt}. 
This is, of course, the more general method being not restricted to our specific Hamiltonian.
The initial value $G(t=0^+)$ is given quite generally by the ``affine shear elasticity" $\muA$,
i.e. the ensemble-averaged second functional derivative of the system Hamiltonian with respect 
to an affine shear strain.\cite{WKC16}
Using eqn~(B8) of \citet{WKC16} this yields $\muA = \kBT \nsp/V$ for a system of ideal springs.
Since $\nsp/V \approx m N/V = \rho=1$ and $\kBT =1$, this implies $G(t=0^+)=\muA \approx 1$ 
in agreement with the MC and GRM data presented in the inset of Fig.~\ref{fig_Gt}.
Using general scaling arguments (not restricted to Gaussian chain statistics)\cite{RubinsteinBook}
it is seen that the shear-stress relaxation 
function should decay as
\begin{equation}
G(t) \approx \kBT \rho/N \ \left(\tauN/t \right)^\gamma \ \exp(-t/\tauN) 
\label{eq_Gtscal}
\end{equation}
with $\gamma \equiv 1/\beta = 1/(1+2\nu)$. The first factor $\kBT \rho/N$ stands for the ideal 
pressure of the chains and the last one for the final exponential cutoff. 
If considered at a short local time $t \approx \taumon \sim N^0$, 
eqn~(\ref{eq_Gtscal}) becomes 
$G(t\approx 0) \approx \kBT \rho$ in agreement with $G(t = 0^+) = \muA$. 
As seen from eqn~(\ref{eq_Gt_GRM}), the final cutoff is dominated by the relaxation of the smallest mode $p=1$
which suggests to set $\tauN = \tau_{p=1}/2$ for the longest shear-stress relaxation time.
The main panel presents again a dimensionless scaling plot.
We trace the rescaled relaxation modulus $y = G(t) / (\kBT \rho/N)$
as a function of $x = t/\tauN$, i.e. we impose a final cutoff $y \approx \exp(-x)$.
The power-law exponent $\gamma$, characterizing the short time behavior, is again a consequence 
of the scaling requirement that $G(t)$ cannot depend on $N$ in this time regime.
As may be seen from Fig.~\ref{fig_Gt}, the exponent $\gamma = 3/5$ for compact objects 
is confirmed over several orders of magnitude from our data while the standard Rouse exponent 
$\gamma=1/2$ for $\nu=1/2$ (dash-dotted line) is clearly ruled out.
That $\gamma=1/\beta$ holds can be also demonstrated by integration of eqn~(\ref{eq_Gt_GRM}) using eqn~(\ref{eq_taup}). 
Paying attention to the $p=1$ mode this leads to $y(x) \approx x^{-\gamma} \exp(-x)$ for large $I$ as indicated 
by the dashed line in the main panel.
The main point we want to make here is merely that it is in our view inconsistent to claim a value 
$\gamma \neq 1/2$ for the shear-stress relaxation modulus and to attempt then to fit $G(t)$ 
and other dynamical properties using a mode expansion in terms of relaxation times $\taup$ 
characterized by an exponent $\beta=2$.\footnote[8]{We remind that an exponent $\gamma \approx 0.4$ 
has been fitted for dense rings.\cite{Rubinstein08b,Rubinstein13} This compares nicely with the exponent 
$\gamma = 3/7$ predicted recently assuming compact rings.\cite{Rubinstein16}
That our trees are characterized by a significantly larger exponent $\gamma=3/5$ is caused, of course, by the faster
dynamics due to the missing topological constraints.}

\subsection{Dynamical structure factor}
\label{dyna_Fqt}
The vibrational or diffusive motion of folded proteins \cite{Granek12} or 
more general natural or synthetic polymer-like structures 
\cite{Goossen14,Goossen15} can be studied experimentally by means of dynamic light scattering or 
neutron spin echo scattering.\cite{DoiEdwardsBook,RubinsteinBook,BenoitBook}
This allows to extract the single (intrachain) chain dynamical structure factor
\begin{equation}
F(q,t) = \frac{1}{N} \sum_{k,l=1}^{N} 
\langle \exp\left[\text{i} \qvec \cdot (\rvec_k(t) - \rvec_l(0)) \right] \rangle
\label{eq_Fqt_def}
\end{equation}
which generalizes the static structure factor $F(q)=F(q,t=0)$ into the time domain.
\begin{figure}[t]
\centerline{\resizebox{0.9\columnwidth}{!}{\includegraphics*{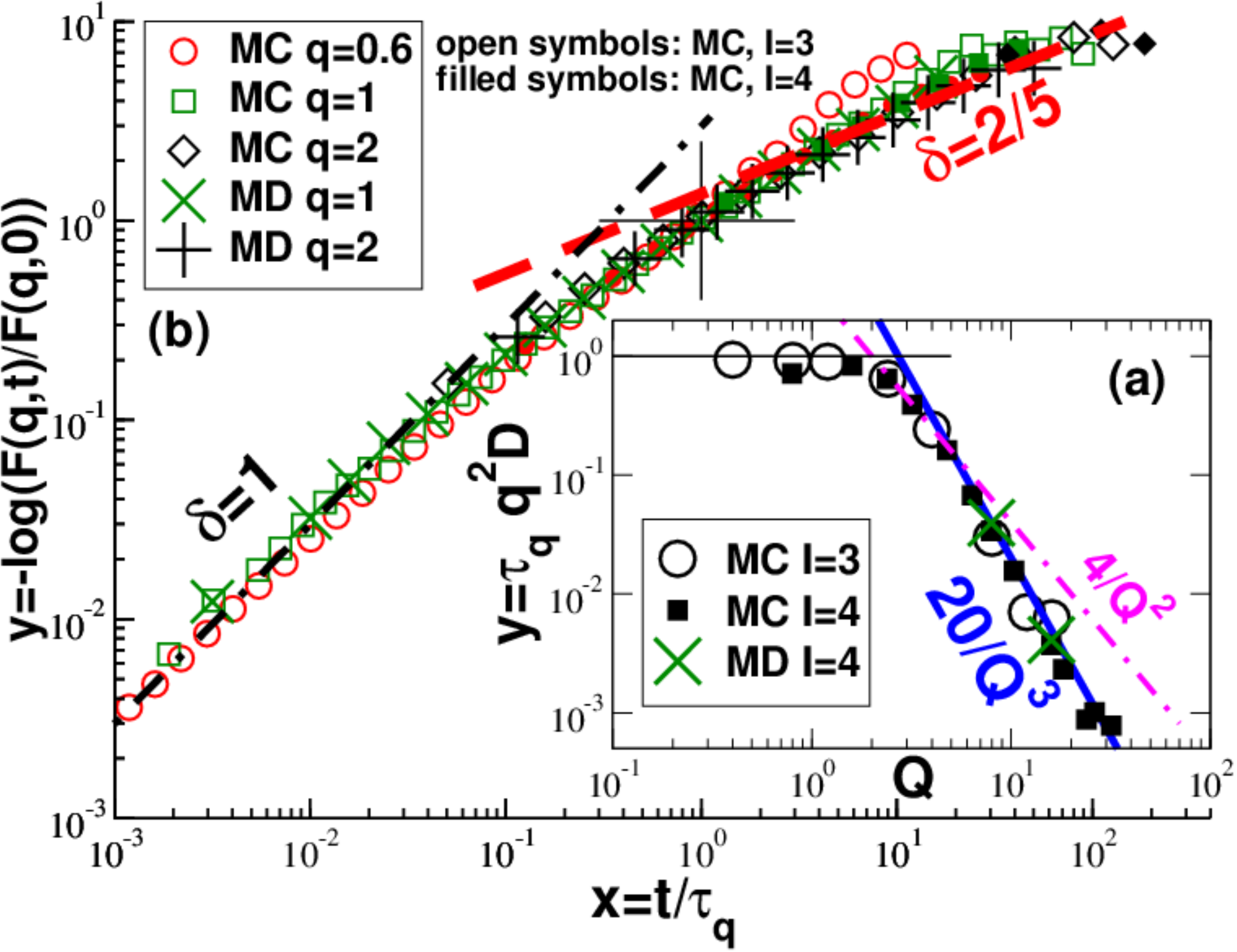}}}
\caption{Dynamical structure factor $F(q,t)$:
{\bf (a)} 
Rescaled relaxation time $y=\tauq q^2 D$ {\em vs.} reduced wavevector $Q$ for MC ($I=3$ and $4$) and MD ($I=4$).  
{\bf (b)}
Reduced dynamical structure factor $y= - \log(F(q,t)/F(q))$ {\em vs.} $x=t/\tauq$.
The dash-dotted line indicates the linear power-law slope $\delta=1$ for small $x$,
the dashed line the slope $\delta=2/5$ for $x \gg 1$. 
}
\label{fig_Fqt}
\end{figure}
In the low-wavevector limit the dynamical structure factor allows to probe quite generally
the overall translational motion of the chain using\cite{DoiEdwardsBook} 
\begin{equation}
F(q,t) \approx N \exp\left(- \frac{q^2}{2d} \ \gcms\right) \ \mbox{ for } Q \equiv q \Rgyr \ll 1.
\label{eq_Fqt_lowq}
\end{equation}
We have verified that this relation holds for both our simulation methods in the low-$Q$ limit (not shown). 

An operationally simple way to quantify the decay of $F(q,t)$ for general wavevectors
is to define a relaxation time $\tauq$ by the time needed for $F(q,t)$ to reach some fraction of its initial value.
We have chosen the standard definition $F(q,t=\tauq)/F(q)=1/\mathrm{e}$.
Following eqn~(\ref{eq_Fqt_lowq}) this implies $\tauq = 1/q^2D$ for $Q \ll 1$.
This suggests the scaling of the reduced relaxation time $y=\tauq q^2 D$ as a function of the reduced wavevector $Q$
shown in panel (a) of Fig.~\ref{fig_Fqt}. Data for $I=3$ and $I=4$ for MC and $I=4$ for MD 
are thus successfully brought to collapse. By construction, $y \to 1$ for $Q \ll 1$
(thin horizontal line). 
The power law for the self-similar $Q$-regime follows from 
\begin{equation}
\tauq \approx \frac{1}{q^2 D} Q^{-\df} \sim N^0/q^{\beta/\nu} 
\mbox{ for } Q \gg 1
\label{eq_tauq_scal}
\end{equation}
with $\beta=1+2\nu$.
As shown by the bold solid line, $y \sim 1/Q^3$ holds as expected for a marginally compact object.
This implies $\tauq \sim 1/q^5$ in the self-similar regime at variance to the standard 
$1/q^4$-scaling.\cite{DoiEdwardsBook}
By setting $1/q \approx b (N/p)^{\nu}$ one confirms using eqn~(\ref{eq_tauq_scal}) that
$\taup \sim (N/p)^{\beta}$ with $\beta=5/3$ (Fig.~\ref{fig_taup}).

Having characterized the relaxation time $\tauq$, we attempt now to
describe the time-dependent decay of $F(q,t)$.
We focus on data corresponding to intermediate times $\taumon \ll t \ll \tauN$ and 
wavevectors $1/\Rgyr \ll q \ll 1/b$ allowing to probe the internal chain motion.
As shown in panel (b) of Fig.~\ref{fig_Fqt}, we plot the reduced dynamical structure factor 
$y = -\log(F(q,t)/F(q))$ as a function of the dimensionless scaling variable $x = t/\tauq$.
Large $y$ corresponds to small density correlations, low $y$ to large ones.
Using a double-logarithmic representation MC and MD data for two chain lengths 
and several wavevectors $q$ are successfully brought to collapse. 
(The MC data for $I=3$ correspond to slightly too small chains for $q=0.6$.)  
Note that we have used the measured relaxation times $\tauq$ for the scaling, 
i.e. by construction $y(x=1)=1$ for all data.
We could have replaced $x=t/\tauq$ by the more common scaling variable $x= t \Gamq$ using the rate \cite{DoiEdwardsBook} 
\begin{equation}
\Gamq = (D/\Rgyr^2) \ Q^{2+\df} \approx 1/\tauq \sim N^0q^5
\label{eq_Gamq}
\end{equation}
in agreement with eqn~(\ref{eq_tauq_scal}) and $\df=3$.
One verifies that the standard Rouse scaling with $\Gamq \sim q^4$ is not appropriate (not shown).

%\paragraph*{Shape of universal function.}
%
We have thus verified the scaling for the intermediate wavevector regime.
The observed universal function is qualitatively similar to the one for linear Rouse chains
for which a steepest decent argument shows that $y$ scales as $\gmon$. 
(See the discussion after eqn~(4.III.10) in ref.~\cite{DoiEdwardsBook}.)
This leads for very small $x$, where the chain connectivity is less important and the
free monomer diffusion is probed, $\gmon \sim t N^0$, to the linear power-law slope ($\delta=1$)
indicated by the dash-dotted line.
The steepest decent argument is also consistent with the fact that we observe for $x \gg 1$ 
a power law $y \sim x^{\delta}$ with the same exponent $\delta=2\nu/\beta=2/5$ (dashed line)
as for $\gmon$ in Fig.~\ref{fig_MSD}.
The data is thus more strongly curved ($\delta=1 \to 2/5$) as for linear chains ($\delta=1 \to 1/2$).

\section{Conclusion}
\label{sec_conc}

\paragraph*{Summary.}
We have investigated theoretically and by means of MC and MD simulations 
several static (Section~\ref{sec_stat}) and dynamical (Section~\ref{sec_dyna}) properties of marginally compact 
hyperbranched polymer trees generated by means of a proper fractal generator (Fig.~\ref{fig_sketch}). 
We have shown that our idealized hyperbranched trees are after at most two iterations of 
the generator self-similar on all scales (Figs.~\ref{fig_ws}-\ref{fig_Fq}), marginally compact 
(Figs.~\ref{fig_contact} and \ref{fig_Fq}) and 
much faster as unbranched objects of same mass (Figs.~\ref{fig_MSD}-\ref{fig_Fqt}).
Two points have been emphasized in the present work.
First, since the return probability decays as 
$p(n) \sim 1/\sqrt{S}n$ for $\none \approx S \ll n \ll \ntwo \approx N$ 
(Fig.~\ref{fig_pn}), the average density of self-contacts $\rcon$ 
must diverge logarithmically, eqn~(\ref{eq_rconfit}), with the total mass $N$,
however, with a prefactor strongly decaying with the spacer length $S$ (Fig.~\ref{fig_contact}).
Second, the standard linear-chain Rouse mode analysis commonly made in experimental studies
\cite{Vlasso16,Goossen14,Goossen15} of the shear-stress relaxation modulus $G(t)$ or 
the intrachain dynamical structure factor $F(q,t)$ 
must necessarily become inappropriate for self-similar compact objects 
(Figs.~\ref{fig_MSD}-\ref{fig_Fqt}) for which the relaxation time of a subchain of mass $n \approx N/p$ 
must scale as $\taup \sim (N/p)^{5/3}$ (Fig.~\ref{fig_taup}).

\paragraph*{Discussion.}
The first emphasized point is the most crucial one of general importance beyond the
polymer-like systems we have focused on.
As correctly stressed by \citet{KK11a}, any description of a real system in terms of a 
marginal-compact model must thus {\em in principle} break down for $N\to \infty$
since for excluded volume constraints the local volume density cannot exceed unity.
However, due to the much stronger dependence on the spacer size (Fig.~\ref{fig_contact}), 
the logarithmic divergence becomes {\em in practice} rapidly irrelevant for a broad mass window. 
Please note that the marginal compact model for melts of rings advanced by some of us 
\cite{MSZ11,obukhovmodel} corresponds to spacer lengths $S$ set by the entanglement length 
$\Nend$ of the corresponding linear chain systems. 
Since $\Nend$ is rather large and considering the ring masses $N$ available at present,
this makes the (certainly simplified) marginal-compact modeling approach at least conceivable. 
Coming back to the living matter network systems mentioned in the Introduction,
it should be a fairly simple task for evolution to tune the length and other structural 
properties of the spacers (twigs, blood vessels, bronchiae, \ldots) connecting the 
metabolically active subunits to increase the allowed mass window $N \ll \Nstar$.
Marginal compact models are thus not only mathematically well-defined
(assuming one or several proper generators) but also of relevance for the description 
of real physical and biological systems. 
In a nutshell, a marginally compact Christmas tree cannot be made only using needles.
One needs twigs as well. 

\paragraph*{Outlook.}
Excluded volume and all other effectively 
long-range interactions have been assumed to be switched off in the present work. 
This is in line with Flory's ideal chain hypothesis for dense polymer melts \cite{DoiEdwardsBook}
and is a necessary condition for taking advantage of the GRM (Appendix~\ref{app_theo}).
However, considering that Flory's hypothesis only holds to leading order for linear chains,\cite{WCX11} 
it may not be justified for more compact fractal objects even for a large spacer length $S$.
Excluded volume effects in melts of hyperbranched chains will thus be considered 
as a function of $S$ in future work.
This should also address the possibility to relax the quenched connectivity matrix $\mathbf{A}$, 
eqn~(\ref{eq_Vspring}), by means of local MC moves allowing small dangling ends 
to hop along the network following broadly the recent work by Rosa and Everaers.\cite{Everaers16a,Everaers16b,Everaers17}

\appendix
\section{Self-contact density $\rcon$}
\label{app_contact}

\paragraph*{Linear chain reference.}
The self-contact density $\rcon$ presented in Fig.~\ref{fig_contact} has been rescaled 
using the self-contact density $\rconlin$ for arbitrarily long linear Gaussian chains. 
Using eqn~(\ref{eq_ws2rcon}) with  $N w(s) = 2$ this reference is
\begin{equation}
\rconlin = 2 \times (d/2\pi b^2)^{d/2} \times \zeta(d/2) \approx 1.724/b^3
\label{eq_rconlin}
\end{equation}
with $\zeta(x) = \sum_{s=1}^{\infty} 1/s^{x}$ being Riemann's $\zeta$-function.\cite{abramowitz}
Note that $\zeta(3/2) \approx 2.612$.
The surprisingly large value of $\rconlin$ stems mainly (by about $50\%$) from the two next and
the two next-nearest neighbors along the chain, i.e. this is a local-range artifact of the extremely 
simplified Gaussian chain model.

\paragraph*{Contributions to $\rcon$.}
Let us trace back eqn~(\ref{eq_rconfit}) to the contributions $\rconone$, $\rcontwo$ and 
$\rconthree$ due to the three regimes of the distribution $w(s)$ discussed in Sec.~\ref{stat_ws}.
For simplicity, we assume large spacer chains and a large iteration number $I$, 
i.e. $1 \ll \sone \approx S \ll \stwo \approx \smax \ll N$. 
The dominant contribution to $\rcon$ thus stems from the second regime ($\sone \ll s \ll \stwo$) 
where $w(s)$ is given by eqn~(\ref{eq_ws_power}).
Integration of eqn~(\ref{eq_ws2rcon}) from $\sone$ to $\stwo$ yields 
\begin{equation}
\frac{\rcontwo}{\rconlin} \approx \frac{\cone}{\sqrt{S}} \times \frac{d}{2} \times \log(\stwo/\sone) 
\label{eq_rcon_two}
\end{equation}
where we have introduced for convenience the constant
\begin{equation}
\cone \equiv \frac{\cws}{d \ \zeta(d/2)} \approx 0.319
\label{eq_cone}
\end{equation}
using that $\cws \approx 5/2$ (Fig.~\ref{fig_ws}).
Since $\sone = S \fone$ and $\stwo = \smax \ftwo$ and using eqn~(\ref{eq_Ntot}) this can be rewritten as
\begin{equation}
\frac{\rcontwo}{\rconlin} \approx \frac{\cone}{\sqrt{S}} \left(\log(N/S) - 
\underline{\frac{d}{2} \log(\fone/\ftwo)} \right)
\label{eq_rcon_twoB}
\end{equation}
where the underlined term in eqn~(\ref{eq_rcon_twoB}) gives one contribution to the 
subdominant correction $\ctwo$ fitted in eqn~(\ref{eq_rcon_twoB}). 
Other subdominant corrections arise from the
linear spacer regime for $1 \le s \ll \sone$ where to leading order
\begin{equation}
\frac{\rconone}{\rconlin} -1 \sim -1/\sqrt{S}.
\label{eq_rcon_oneB}
\end{equation}
The contribution $\rconthree$ of the third regime beyond $\stwo$ is difficult to describe 
by an analytic formula. It is expected to be negligible, $\rconthree \ll \rcontwo$,
due to the cutoff of $w(s)$ seen in Fig.~\ref{fig_ws}. In any case neither $\rconone$
nor $\rconthree$ do matter in the large-$I$ limit. 
Summarizing all three terms this leads to eqn~(\ref{eq_rconfit})
where we have lumped all subdominant contributions into the coefficient $\ctwo$.
\section{Additional points concerning $F(q)$}
\label{app_Fq}

\paragraph*{Spherical preaveraging.}
As reminded at the beginning of Section.~\ref{stat_Fq}, $F(q)$ is the
ensemble average of the squared Fourier transform $\rhohat(\qvec)$
of the fluctuating instantaneous density $\sum_k \delta(\rvec-\rvec_k)$.
Following \citet{Likos03,Likos05}
this begs the question of whether in the limit of large marginally compact chains
density fluctuations become sufficiently small allowing to replace $\rhohat(\qvec)$
by the Fourier transform of the averaged density profile $\rho_{k=1}(r)$ around the root monomer.
(Per symmetry the root monomer stands at the center of the preaveraged density profile.)
The (not shown) density $\rho_{k=1}(r)$ is very similar to the pair correlation
density $\rpair(r)$ averaged over all monomers $k$ (Fig.~\ref{fig_pair}).
It may be obtained from the simulations or by means of
\begin{equation}
\rho_{k=1}(\rvec) = N \sum_{s=0}^{\smax} \wroots P(\rvec,s)
\label{eq_pr_wroots}
\end{equation}
with $P(\rvec,s)$ being given by eq.~(\ref{eq_Grsgauss}).
Due to the spherical symmetry this suggests using eqn~(6.54) of ref.~\cite{BenoitBook} that
\begin{equation}
F(q) \approx \frac{1}{N} \ \left( \int \ddiff\rvec \ \rho_{k=1}(r) \ \frac{\sin(\qvec \cdot \rvec)}{\qvec \cdot \rvec} \right)^2.
\label{eq_Fq_preaver}
\end{equation}
Note that this is equivalent to the factorization suggested on page S1791 by \citet{Likos05}.
As revealed by the dash-dotted line in Fig.~\ref{fig_Fq},
eqn~(\ref{eq_Fq_preaver}) is {\em not} a useful approximation.
This counter example shows that it is not possible in general to determine from the measured form factor $F(q)$
the average density profile and thus an effective interaction potential. While this approach may work
for colloids, collapsed polymer chains below the $\theta$-point or densely packed dendrimers
\cite{Likos03,Likos05} having a clearly defined surface, it should be used with care for general
compact objects. Compactness does not necessarily imply negligible density fluctuations.

\begin{figure}[t]
\centerline{\resizebox{0.9\columnwidth}{!}{\includegraphics*{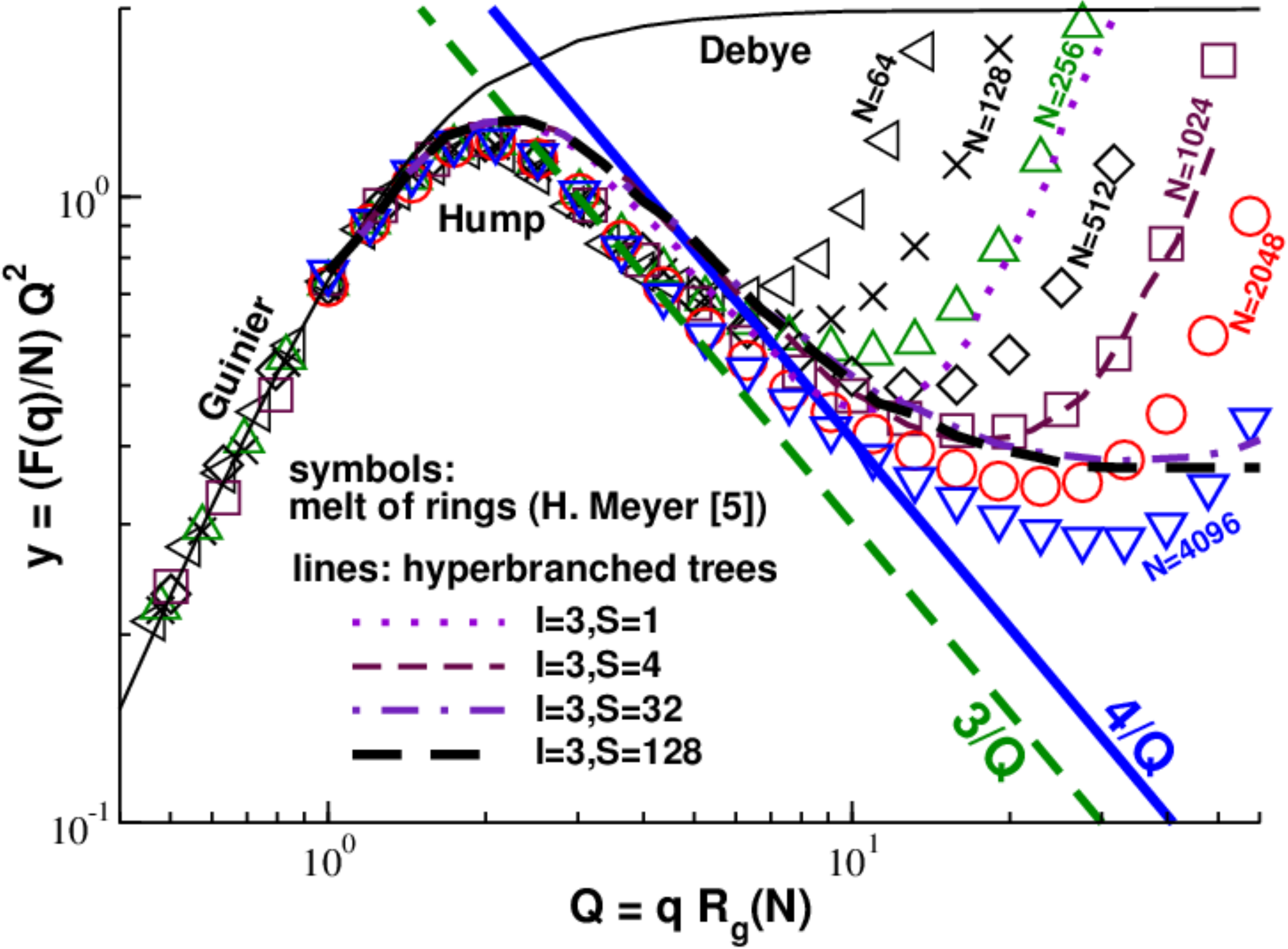}}}
%\vspace*{-1.5cm}
\caption{Kratky representation $y=(F(q)/N) Q^2$ {\em vs.} $Q=q\Rgyr$ comparing 
our simple tree model for $I=3$ and different $S$ as indicated with published data 
for ring melts obtained by MD simulations of a standard bead-spring model.\cite{MSZ11}
Both data sets are qualitatively similar, albeit with slightly different prefactors for the $1/Q$-slopes.
The thin solid line indicates the Debye formula for linear Gaussian chains.\cite{DoiEdwardsBook}
}
\label{fig_ring_Fq}
\end{figure}

\paragraph*{Spacer length effects.}
The data presented in Fig.~\ref{fig_Fq} have been obtained for spacer chains of length $S=1$.
The observed strong increase of the data for large $q$ is simply due to the discrete monomeric units used 
in our simulations. As shown in Fig.~\ref{fig_ring_Fq} for $I=3$, a horizontal 
plateau gradually appears with increasing $S$ corresponding to the density fluctuations associated to the 
Gaussian spacer chain. This plateau is roughly located at the minimum $y_{min} \sim \Rgyr^2/N \sim 1/N^{1/3}$ 
of the data for $S=1$. It is thus not possible by tuning the local physics, i.e. the spacer length $S$ and
(as may be shown) the statistical segment size $b$, to increase the range of the intermediate wavevector regime.
Since $y_{min} \sim 1/N^{1/3}$ decreases extremely weakly with mass, the determination of the $1/Q$-scaling 
becomes very difficult.  

\paragraph*{Comparison with ring data.}
This sluggish convergence is by no means unique to our simple model, but is generic 
for more-or-less compact polymers such as linear polymer melts in strictly two dimensions 
\cite{MKA09,MWK10,MSZ11} or melts of polymer rings.\cite{MSZ11,WMJ13,Goossen14,Goossen15}
This may be seen from the ring data\cite{MSZ11} also included in Fig.~\ref{fig_ring_Fq}. 
As expected from the sluggish convergence of the radius of gyration, 
the reduced ring data approach from {\em above} the slope $y=3/Q$ (bold dashed line) indicated to guide the eye.
The ring simulation data are thus {\em not} compatible with a generalized Porod law, eqn~(\ref{eq_GPL}), 
with a fractal surface dimension $\ds < d$. 
Similar behavior is also seen in other numerical studies \cite{MWC96,MWC00,KK11a,WMJ13} and 
in recent experimental work.\cite{Goossen14,Goossen15} 
As emphasized by \citet{WMJ13}, it is important that the data is correctly scaled used 
the measured radius of gyration as done in Fig.~\ref{fig_ring_Fq}.
Interestingly, albeit our tree model and the ring data are qualitatively similar,
the respective scaling functions $f(Q)$ of both systems differ slightly: 
The trees reveal stronger density fluctuations with a broader hump and a slightly 
larger prefactor for the $1/Q$ power law. 
While for large wavevectors the fit of the tree model onto the ring data can be improved
by tuning $S$ and $b$, this is impossible in the $Q$-scaling regime. 
It is currently not clear whether it is possible to generalize the tree generator, 
e.g., using a multi-fractal approach,\cite{obukhovmodel}
to get a better match of the scaling functions $f(Q)$. 

\section{Generalized Rouse Model (GRM)}
\label{app_theo}

The dynamics of our ideal spring networks, eqn~(\ref{eq_Vspring}),
is described by the Langevin equation \cite{DoiEdwardsBook}
\begin{equation}
\zeta \rvecdot_i+\partial_{\rvec_i} V(\{\rvec_i\})= \fvec_i(t),
\label{eq_app_langevin}
\end{equation}
where the stochastic forces $\underline{f}_i(t)$ are represented by a white noise and 
$\zeta$ is the friction constant. 
Due to the bilinear form of $V(\{\rvec_i\})$, eqn~(\ref{eq_Vspring}), the set of Langevin equations 
(eqn~\ref{eq_app_langevin}) is a linear set. It is solved by diagonalization of $\mathbf{A}$.
The non-vanishing ($p\ge 1$) eigenvalues $\lamp$ yield the relaxation times 
$\taup \equiv \taumon /\lamp$ with $\taumon=\zeta/K$. 
Using merely the eigenvalues $\lamp$ 
of the connectivity matrix $\mathbf{A}$ (and not its eigenvectors), 
many dynamic properties can be readily calculated \cite{schiessel98,gurtovenko05} 
as shown by the relations given in the main text for the monomer MSD, eqn~(\ref{eq_gmon_GRM}), 
and the shear-stress relaxation modulus $G(t)$, eqn~(\ref{eq_Gt_GRM}).
Unfortunately, the computation of the dynamical structure factor $F(q,t)$
requires in addition the determination of all eigenvectors.

In the reminder we corroborate the scaling relations presented in Section~\ref{sec_dyna}
using a more general context stemming from the analysis of the eigenmode spectrum.\cite{Alexander82}
For fractals, the density of states $h(\lambda)$ usually possesses a power law behavior \cite{Alexander82}
\begin{equation}\label{eq_app_dens_states}
h(\lambda) \equiv \frac{1}{N-1} \sum_{p=1}^{N-1} \delta(\lambda - \lamp) 
\sim\lambda^{\frac{d_s}{2}-1},
\end{equation}
where $d_s$ is the so-called \textit{spectral dimension}.\cite{Alexander82}
Equation~(\ref{eq_app_dens_states}) stems from solid state physics, 
where for regular lattices one gets a similar expression having instead of 
$d_s$ the dimension of the lattice.\cite{ziman72} However, in case of fractals, 
the spectral dimension $d_s$ differs generally from the dimension of the fractal lattice.\cite{Alexander82}
We note also that the density of states 
$h(\lambda)$ is directly connected to the dependency of the eigenvalues $\lambda_p$ on the mode number $p$. 
Indeed, rewriting $p=p(\lambda)$ and bearing in mind that $h(\lambda)\sim\partial_{\lambda}p(\lambda)$, 
one gets $\lamp \sim(p/N)^{2/d_s}$ and thus 
\begin{equation}
\taup \approx \taumon (N/p)^{\beta} \mbox{ with } \beta = 2/d_s.
\label{eq_app_taup}
\end{equation}
Hence, the smallest non-vanishing eigenvalue $\lambda_{\min}$ and the
corresponding largest relaxation time $\tauN$ are 
\begin{equation}
\lambda_{\min}\sim N^{-2/d_s} \mbox{ and } \tauN \approx \taumon N^{2/d_s}.
\label{eq_lammin}
\end{equation}
Moreover, in many cases the spectral dimension $d_s$ is related to the fractal dimension $\df$ of a Gaussian macromolecule 
by \cite{Alexander82,Cates84}
\begin{equation}
\label{eq_app_d_s_d_f}
d_s=\frac{2\df}{\df+2} \mbox{ hence, } \beta = 1 + 2/\df.
\end{equation}
To see this relation, one considers the continuous limit 
$N \Rgyr^2 \sim \ \sum(1/\lambda)\rightarrow\int\mathrm{d}\lambda(h(\lambda)/\lambda)$.\cite{forsman76}
This leads readily to $\Rgyr^2 \sim N^{(2-d_s)/d_s}$,
if one takes eqn~(\ref{eq_lammin}) into account.\cite{sommer95}  
It is important to note, that there are examples, where the relation of 
eqn~(\ref{eq_app_d_s_d_f}) is invalid, partly due to violation of the $\lambda_{\min}\sim N^{-2/d_s}$ behavior, see, e.g., 
\citet{sokolov16}. Now, the scaling of eqn~(\ref{eq_app_dens_states}) determines the behavior of the dynamic variables. 
One finds\cite{schiessel98,blumen04} for intermediate times $\taumon \ll t \ll \tauN$ 
\begin{equation}
\gmon \sim t^{1-\frac{d_s}{2}} \textrm{ and } G(t)\sim t^{-\frac{d_s}{2}}.
\end{equation}
Since $\df=3$, we have $d_s=6/5$. Hence, from the above relations we get
$\taup \sim (N/p)^{5/3}$, $\gmon \sim t^{2/5}$ and $G(t)\sim t^{-3/5}$
in agreement with the corresponding statements made in Section~\ref{sec_dyna}.

%\balance

%\renewcommand\refname{Notes and references}

%%%REFERENCES%%%
\bibliographystyle{rsc} %the RSC's .bst file
%\bibliography{../../bibl/BOOK,../../bibl/ANS,../../bibl/Biswas,../../bibl/Blumen,../../bibl/Alexander,../../bibl/Borisov,../../bibl/Burchard,../../bibl/Cates,../../bibl/Davies,../../bibl/Dolgushev,../../bibl/deGennes,../../bibl/Duplantier,../../bibl/Everaers,../../bibl/Grosberg,../../bibl/Granek,../../bibl/KK,../../bibl/Likos,../../bibl/JPW,../../bibl/Mirny,../../bibl/Nechaev,../../bibl/Obukhov,../../bibl/Richter,../../bibl/Rubinstein,../../bibl/Stockmayer,../../bibl/Turner,../../bibl/Vlasso,../../bibl/Wiener,../../bibl/Winkler,bibl,../dend,../../bibl/ring,../../bibl/LAMMPS}

\providecommand*{\mcitethebibliography}{\thebibliography}
\csname @ifundefined\endcsname{endmcitethebibliography}
{\let\endmcitethebibliography\endthebibliography}{}

\end{document}